\documentclass[aps,prc,twocolumn,10pt,superscriptaddress]{revtex4-1}

\usepackage{csquotes} 
\usepackage{hyperref}
\usepackage{graphicx}
\usepackage{amsmath}
\usepackage[usenames,dvipsnames,svgnames,table]{xcolor}
\usepackage{slashed} 
\usepackage{physics} 

\hypersetup{
  colorlinks,
  unicode=True,
  linkcolor=Blue,
  citecolor=Blue,
  urlcolor=Blue,
  pdftitle={Inelastic and elastic parton scatterings in the strongly interacting quark-gluon plasma},
  pdfauthor={Ilia Grishmanovskii},
  pdfsubject={HEP}
}

\graphicspath{ {figures/} }

\newcommand{\ubar}{\bar{u}}
\newcommand{\dbar}{\bar{d}}
\newcommand{\sbar}{\bar{s}}
\newcommand{\srs}{\sqrt{s}}
\newcommand{\dpi}{(2\pi)}
\newcommand{\GL}[3]{\varepsilon^{#1}_{#2}(#3)}
\newcommand{\GP}[3]{\Pi_{#1#2}(#3)}

\newcommand{\VX}[3]{V^{#1,#2}_{#3}}
\newcommand{\QP}[1]{\Lambda(#1)}
\newcommand{\GV}[3]{\Omega^{#1,#2}(#3)}

\newcommand{\sld}[1]{\slashed{#1}}
\newcommand{\Msq}[1]{|\bar{\mathcal{M}}_{#1}|^2}

\begin{document}


\title{Inelastic and elastic parton scatterings in the strongly interacting quark-gluon plasma}


\author{Ilia Grishmanovskii}
\email{grishm@itp.uni-frankfurt.de}
\affiliation{Institut f\"ur Theoretische Physik, Johann Wolfgang Goethe-Universit\"at,Max-von-Laue-Straße 1, D-60438 Frankfurt am Main, Germany}

\author{Olga Soloveva}
\affiliation{Helmholtz Research Academy Hesse for FAIR (HFHF), GSI Helmholtz Center for Heavy Ion Physics, Campus Frankfurt, 60438 Frankfurt, Germany}
\affiliation{Institut f\"ur Theoretische Physik, Johann Wolfgang Goethe-Universit\"at,Max-von-Laue-Straße 1, D-60438 Frankfurt am Main, Germany}

\author{Taesoo Song}
\affiliation{GSI Helmholtzzentrum f\"ur Schwerionenforschung GmbH,Planckstrasse 1, D-64291 Darmstadt, Germany}

\author{Carsten Greiner}
\affiliation{Institut f\"ur Theoretische Physik, Johann Wolfgang Goethe-Universit\"at,Max-von-Laue-Straße 1, D-60438 Frankfurt am Main, Germany}
\affiliation{Helmholtz Research Academy Hesse for FAIR (HFHF), GSI Helmholtz Center for Heavy Ion Physics, Campus Frankfurt, 60438 Frankfurt, Germany}

\author{Elena Bratkovskaya}
\affiliation{GSI Helmholtzzentrum f\"ur Schwerionenforschung GmbH,Planckstrasse 1, D-64291 Darmstadt, Germany}
\affiliation{Institut f\"ur Theoretische Physik, Johann Wolfgang Goethe-Universit\"at,Max-von-Laue-Straße 1, D-60438 Frankfurt am Main, Germany}
\affiliation{Helmholtz Research Academy Hesse for FAIR (HFHF), GSI Helmholtz Center for Heavy Ion Physics, Campus Frankfurt, 60438 Frankfurt, Germany}


\date{\today}
\begin{abstract}
We investigate the role of inelastic processes in the strongly interacting quark-gluon plasma (sQGP) based on the effective dynamical quasiparticle model (DQPM). In the DQPM the nonperturbative properties of the sQGP at finite temperature $T$ and baryon chemical potential $\mu_B$ are described in terms of strongly interacting off-shell partons (quarks and gluons) with dynamically generated spectral functions whose properties are adjusted to reproduce the lattice QCD equation of state for the quark-gluon plasma in thermodynamic equilibrium. For the first time the massive gluon radiation processes from the off-shell quark-quark ($q+q$) and quark-gluon ($q+g$) scatterings are calculated explicitly within leading order Feynman diagrams with effective propagators and vertices from the DQPM without any further approximations. We present the results for the energy and temperature dependencies of the total and differential radiative cross sections and compare them to the corresponding elastic cross sections. We show that our results reproduce the perturbative QCD calculations in the limit of zero masses and widths of quasiparticles. Also we study the $\mu_B$ dependence of the inelastic cross sections. Moreover, we present estimates for the transition rate and relaxation time of radiative versus elastic scatterings in the sQGP.
\end{abstract}

\maketitle


\section{Introduction}

Heavy-ion collisions provide a unique possibility to explore the structure of the QCD phase diagram and study the properties of the quark-gluon plasma (QGP) under extreme conditions. Experimental observations \cite{BRAHMS:2004adc,PHENIX:2004vcz,PHOBOS:2004zne,STAR:2005gfr} suggest that the matter created in central heavy-ion collisions shows the properties of a strongly coupled liquid \cite{Shuryak:2004cy,Gyulassy:2004zy} rather than a weakly coupled gas of quark and gluons as expected initially. The theoretical description of such a strongly interacting QGP is a challenging task since perturbative methods of QCD do not work in the vicinity of the critical temperature $T_c$ where the strong coupling constant is large \cite{Ghiglieri:2018dib}. Moreover, a solution of QCD on the lattice (lQCD) can provide presently only a limited knowledge about the phase diagram at moderate baryon $\mu_B$ (or quark $\mu_q=\mu_B/3$) chemical potential where $\mu_B \simeq 3\mu_q$.

To overcome these difficulties and obtain information about the degrees-of-freedom of the strongly interacting QGP (sQGP) and their properties effective models are applied such as quasiparticle models (QPMs) \cite{Blaizot:2000fc,Peshier:1995ty,Sambataro:2023zkr,Liu:2021dpm}. One of the prominent quasiparticle models is the QCD-inspired effective field-theoretical Dynamical QuasiParticle Model (DQPM) introduced in Ref. \cite{Peshier:2005pp} and worked out in Refs. \cite{Cassing:2007nb,Cassing:2007yg,Berrehrah:2016vzw,Moreau:2019vhw,Soloveva:2019xph} and others. The main difference and advantage of the DQPM with respect to other QPMs is that it is by construction a two particle irreducible (2PI) model while the other quasiparticle models are one particle irreducible (1PI) in nature. The DQPM is based on a propagator representation with effective (anti)quarks and gluons which properties are defined by complex self-energies. Thus, in addition to the temperature (and $\mu_B$ in DQPM) dependence of quasiparticle masses -- as in other effective QPMs -- the DQPM contains the information about their interactions in the sQGP via dynamically generated widths, i.e., the imaginary part of self-energies, which also depend on temperature $T$ (and $\mu_B$). Consequently, the DQPM quasiparticles are off-shell broad states with dynamically generated spectral functions whose properties (masses and widths) are adjusted by fitting the DQPM entropy density to the lQCD data at zero $\mu_B$ \cite{Cassing:2007nb,Cassing:2007yg}. This avoids an introduction of a Debye mass in the propagator for the regularization of infrared divergences when calculating cross sections as well as an extra bag constant when computing the thermodynamic properties of the QGP.

The effective off-shell strongly interacting partonic quasiparticles of the DQPM allow for a microscopic interpretation of the lattice QCD results in the $T,\mu_B$ plane for thermodynamic quantities -- such as entropy density, pressure, and energy density -- as well as for transport coefficients which measure the interactions of the QGP degrees-of-freedom in thermodynamic equilibrium. The transport coefficients are evaluated by computing the off-shell partonic elastic interaction cross sections based on the leading-order scattering Feynman diagrams with effective propagators and vertices and depend on $T, \mu_B$, the invariant energy of the colliding partons, $\sqrt{s}$, as well as the scattering angle \cite{Moreau:2019vhw}. The DQPM results for the shear and bulk viscosity, electric conductivity, magnetic susceptibility, as well as the full diffusion matrix show a good agreement with existing lQCD results as well as with a Bayesian analysis based on the hydrodynamic description of the heavy-ion data \cite{Ozvenchuk:2012kh,Cassing:2013iz,Steinert:2013fza,Moreau:2019vhw,Fotakis:2021diq,Soloveva:2020hpr}. We note that the DQPM quasiparticles have been incorporated in the Parton Hadron String Dynamics (PHSD) approach, based on Kadanoff-Baym off-shell dynamics, for the description of the partonic phase and hadronization (cf. Refs. \cite{Cassing:2008nn,Linnyk:2015rco,Bleicher:2022kcu}), which has led to a successful description of a large variety of experimental data.

Recently, in Ref. \cite{Grishmanovskii:2022tpb} the DQPM was applied for studying the elastic interactions of leading jet partons in the sQGP and for the calculation of the jet transport coefficients, i.e. the transverse momentum transfer squared per unit length $\hat q$ as well as the energy loss per unit length $\Delta E =dE/dx$. However, for a realistic description of jet attenuation in heavy-ion collisions, in addition to elastic scattering, the gluon radiation processes have to be accounted for, for which contribution grows with increasing energy of jet partons \cite{Djordjevic:2006tw,Stojku:2020tuk}. The role of radiative processes for jet suppression has been extensively studied by a variety of groups (cf. Refs. \cite{Cao:2020wlm} and reference therein). Moreover, the radiative processes are also important for an understanding of the energy loss of heavy quarks in heavy-ion collisions: while at low and intermediate $p_T$ the elastic processes are dominant, at high $p_T$ the gluon radiation can be sizable \cite{Cao:2018ews}.

For the case of the thermalized strongly interacting QGP -- as realized in the DQPM -- one expects that the inelastic processes of the gluon radiation with large thermal masses from quark-quark ($q+q$) and quark-gluon ($q+g$) scatterings ($q+q \to q+q +g$ and $q+g \to q+g+g$) should be strongly suppressed relative to the elastic scattering since the average collision energy of thermal partons is low and not sufficient for the emission of heavy gluons. Consequently, in the sQGP -- consisting of thermal quasiparticles -- such radiative processes are expected to be even more suppressed relative to the pQCD plasma due to the large mass of the emitted gluons and final off-shell partons. However, such an expectation has to be quantified which is a motivation for this study.

Thus, the goal of this study is to extend the DQPM to radiative processes for the $q+q$ and $q+g$ scatterings, to calculate the inelastic $q+q \to q+q +g$ and $q+g \to q+g+g$ cross sections, to estimate the inelastic rate relative to the elastic one, and to find the energy and temperature regimes where radiative processes become visible or even dominant in the strongly interacting QGP. The main difference from previous studies is that now we avoid any kind of approximations (on the gluon energy, transverse momentum transfer, etc.) and calculate all possible $2 \to 3$ diagrams explicitly. This approach will allow us to explore the entire radiative processes from massive partons with spectral functions and study different limits and applications.

We recall that the study of radiative processes in QCD has a long history. Explicit analytical expressions for the $2\to 3$ amplitudes in the case of massless pQCD quarks were derived in Refs. \cite{Berends:1981rb,Ellis:1985er}. However, the actual calculations of cross sections for radiative processes in pQCD with massless quarks and gluons require to regulate an infrared divergence of the amplitudes realized by the introduction of a Debye mass (which is a model parameter) in propagators. Moreover, big efforts were related to account for the Landau-Pomeranchuk-Migdal (LPM) effect \cite{Landau:1953um,Migdal:1956tc}, which suppresses the soft gluon radiation \cite{Wang:1994fx,Zakharov:1996fv,Zakharov:2020whb}. In recent years, there has been a renewed focus on improving the accuracy of analytical results. It includes efforts to go beyond the limitations of soft-emissions approximations, as highlighted in Refs. \cite{Arnold:2020uzm,Arnold:2021pin}. Additionally, there has been significant progress in the resummation of multiple scatterings, a topic extensively reviewed and discussed in recent studies \cite{Andres:2022ndd,Andres:2023jao,Barata:2020sav}.

To simplify the calculations, different approximations are used. A commonly used approximation to the leading order pQCD matrix element for partonic $2 \to 3$ processes is a formula derived by Gunion and Bertsch (GB) in 1981 \cite{Gunion:1981qs}. The original GB approximation is strictly valid at midrapidity and implies that the radiated gluon and the momentum transfer of the process are soft and all particles are massless. In Refs. \cite{Das:2010hs,Abir:2010kc,Bhattacharyya:2011vy} efforts are made to go beyond the soft GB approximation while still obtaining a relatively compact form. In Ref. \cite{Fochler:2013epa} the GB approximation was extended to be valid in the entire region of rapidity and in Ref. \cite{Uphoff:2014gaa} the latter was generalized to the case of heavy quarks. These approximations have been widely used when solving transport problems, for example via rate equations or via microscopic transport approaches, most notably the study of heavy quark and jet suppression within the microscopic transport model Boltzmann Approach to MultiParton Scatterings (BAMPS) \cite{Uphoff:2014cba,Senzel:2020tgf,Fochler:2013epa,Senzel:2020tgf}.

The influence of radiative processes on jet attenuation beyond pQCD has been studied in the framework of finite temperature field theory in the hard thermal loop (HTL) approximation \cite{Djordjevic:2006tw,Djordjevic:2008iz,Djordjevic:2009cr,Djordjevic:2011dd,Caron-Huot:2008zna}, in nonperturbative calculations within electrostatic QCD (EQCD) \cite{Moore:2019lgw,Moore:2021jwe} and in anti–de Sitter/conformal field theory (AdS/CFT) models \cite{Liu:2006ug, Grefa:2022sav}, which show the importance of nonperturbative effects for an understanding of the transverse jet momentum broadening and jet quenching \cite{Moore:2021myb,Schlichting:2021idr,Andrews:2018jcm,Casalderrey-Solana:2020jbx,Shi:2022rja}.

The emission of massless and massive gluons by heavy quark scattering with massless quarks and gluons from the QGP medium has been studied within scalar pQCD in Refs. \cite{Gossiaux:2010yx,Aichelin:2013mra,Gossiaux:2014jga}. It has been shown that the emission rate decreases with increasing gluon mass.

Recently, in Ref. \cite{Song:2022wil} the soft gluon approximation (SGA) has been applied to evaluate the massive gluon emission from the nonperturbative sQGP medium described by the DQPM. The SGA simplifies the calculations, making the scattering amplitude factorizable into the elastic scattering and the emission of a soft gluon. In the present study, we depart from this approximation which makes our results applicable for the emission of gluons of any energy and angular direction. However, we mention (without showing explicitly) that the SGA qualitatively agrees with our exact results for gluon radiation cross sections in the range of validity of the SGA.

The paper is organized as follows: We start with the Introduction in Sec. I and we recall the basis of the DQPM model in Sec. II. In Sec. III we present the framework for the calculation of radiative processes from quark-quark and quark-gluon scattering in the DQPM. In Sec. IV we continue with the results for the inelastic and elastic total and differential cross sections as a function of temperature and scattering energy, transition rates and relaxation times for quarks and gluons in the sQGP. We finish the paper with a summary in Sec. V. Moreover, the details of our calculations for the cross sections are presented in the appendixes.


\section{Dynamical Quasiparticle Model}
\label{sec:DQPM}

We start with a brief reminder of the basis of the DQPM and refer the reader to Ref. \cite{Soloveva:2020hpr} for a more detailed description of the latest version of the DQPM.

The Dynamical Quasiparticle Model (DQPM) \cite{Peshier:2005pp,Cassing:2007nb,Cassing:2007yg,Berrehrah:2016vzw,Moreau:2019vhw,Soloveva:2020hpr} is an effective model which describes the QGP in terms of strongly interacting quarks and gluons. This model is based on fitting the properties of these particles in order to reproduce the results of lattice QCD calculations in thermal equilibrium and at vanishing chemical potential. 

\noindent
The quasiparticles in the DQPM are characterized by the following properties: \\
$\bullet$ \textquote{Dressed} propagators, i.e., single-particle (two-point) Green's functions, have the form
\begin{equation}
  G^{R}_j (\omega, {\bf p}) = \frac{1}{\omega^2 - {\bf p}^2 - M_j^2 + 2 i \gamma_j \omega}
\label{eq:propdqpm}
\end{equation}
for quarks, antiquarks, and gluons ($j = q,\bar q,g$), using $\omega=p_0$ for energy, the widths $\gamma_{j}$ and the masses $M_{j}$. \\
$\bullet$ The model uses complex self-energies for gluons, $\Pi = M_g^2-2i \omega \gamma_g$, and for (anti)quarks, $\Sigma_{q} = M_{q}^2 - 2 i \omega \gamma_{q}$, where the real part of the self-energies is associated with dynamically generated thermal masses, while the imaginary part provides information about the lifetime and reaction rates of the particles.

The spectral functions in the DQPM are no longer $\delta$ functions, but have a finite width $\gamma_{j}$ \cite{Linnyk:2015rco}:
\begin{align}
\label{eq:spectral_function}
  \rho_{j}(\omega,{\bf p}) &= \frac{\gamma_{j}}{\tilde{E}_j}
  \left(\frac{1}{(\omega-\tilde{E}_j)^2+\gamma^{2}_{j}}
  -\frac{1}{(\omega+\tilde{E}_j)^2+\gamma^{2}_{j}}\right) 
  \nonumber\\
  &\equiv \frac{4\omega\gamma_j}{\left( \omega^2 - \vec{p}^2 - M^2_j \right)^2 + 4\gamma^2_j \omega^2}
\end{align}
Here, $\tilde{E}_{j}^2({\bf p})={\bf p}^2+M_{j}^{2}-\gamma_{j}^{2}$. The spectral function is antisymmetric in $\omega$ and normalized as
\begin{equation}
\label{eq:spectral_function_norm}
  \int\limits_{-\infty}^{\infty}\frac{d\omega}{2\pi}\
  \omega \ \rho_{j}(\omega,{\bf p})=
  \int\limits_{0}^{\infty} d\omega \frac{\omega}{\pi}\ 
  \rho_{j}(\omega,{\bf p})=1.
\end{equation}

$\bullet$ A model ansatz is used for the masses $M_{j}(T,\mu_q)$ and widths $\gamma_{j}(T,\mu_q)$ as functions of the temperature $T$ and the quark chemical potential $\mu_q$.

With the quasiparticle properties (or propagators) fixed as described above, one can evaluate thermodynamic quantities such as the entropy density $s(T,\mu_B)$, the pressure $P(T,\mu_B)$, and energy density $\epsilon(T,\mu_B)$ in a straightforward manner by starting with the entropy density and number density in the propagator representation from Baym \cite{Vanderheyden:1998ph,Blaizot:2000fc}.

By comparison of the entropy density -- computed within the DQPM framework -- to the lQCD data, one can fix the few parameters used in the ansatz for quasiparticle masses and widths.


\subsection{Quasiparticle properties}

The following ansatz is used in the DQPM for the definition of the quasiparticle properties (masses and widths) as functions of $T$ and $\mu_q$:\\
$\bullet$ The dynamical quasiparticle \textit{pole masses} are given by the HTL thermal mass in the asymptotic high-momentum regime, i.e., for gluons by \cite{Bellac:2011kqa,Linnyk:2015rco}
\begin{equation}
\label{eq:Mg}
  M^2_{g}(T,\mu_q)=\frac{g^2(T,\mu_q)}{6}\left(\left(N_{c}+\frac{1}{2}N_{f}\right)T^2
  +\frac{N_c}{2}\sum_{q}\frac{\mu^{2}_{q}}{\pi^2}\right),
\end{equation}
and for quarks (antiquarks) by
\begin{equation}
\label{eq:Mq}
  M^2_{q(\bar q)}(T,\mu_q)=\frac{N^{2}_{c}-1}{8N_{c}}g^2(T,\mu_q)\left(T^2+\frac{\mu^{2}_{q}}{\pi^2}\right),
\end{equation}
where $N_{c}\ (=3)$ stands for the number of colors and $N_{f}\ (=3)$ denotes the number of flavors. Equation \eqref{eq:Mq} determines masses for the ($u,d$) quarks; the strange quark has a larger bare mass for controlling the strangeness ratio in the QGP. Empirically, we find $M_s(T,\mu_B)= M_{u/d}(T,\mu_B)+ \Delta M$, where $\Delta M \simeq 30$ MeV, which has been fixed once in comparison to experimental data \cite{Moreau:2019vhw}. 

$\bullet$ The effective quarks, antiquarks, and gluons in the DQPM acquire sizable \textit{widths} $\gamma_j$, which are taken in the form \cite{Linnyk:2015rco}
\begin{equation}
\label{eq:widths}
	\gamma_{j}(T,\mu_\mathrm{B}) = \frac{1}{3} C_j \frac{g^2(T,\mu_\mathrm{B})T}{8\pi}\ln\left(\frac{2c_m}{g^2(T,\mu_\mathrm{B})}+1\right).
\end{equation}
Here $c_m=14.4$ is related to a magnetic cutoff, which is an additional parameter of the DQPM, and $C_q = \dfrac{N_c^2 - 1}{2 N_c} = 4/3$ and $C_g = N_c = 3$ are the QCD color factors for quarks and for gluons, respectively. We also assume that all (anti)quarks have the same width. 

$\bullet$ As follows from Eqs. (\ref{eq:Mq}) and (\ref{eq:widths}), the masses and widths depend on the \textit{coupling constant} (squared), $g^2 = 4\pi\alpha_s$, which defines the strength of the interaction. This parameter is a critical quantity that determines all the microscopic properties and transport coefficients of the DQPM, including the thermal masses and widths. 

In the DQPM, the value of $g^2$ is extracted from lQCD by utilizing a parametrization method introduced in Ref. \cite{Berrehrah:2015vhe}, where it has been shown that for a given value of $g^2$, the ratio $s(T,g^2)/T^3$ is almost constant for different temperatures, i.e., ${\frac{\partial}{\partial T}} (s(T,g^2)/T^3)=0$. Therefore, the entropy density $s$ and the dimensionless equation of state in the DQPM is a function of the effective coupling only, i.e., $s(T,g^2)/s_{SB}(T) = f(g^2)$, where $s^{QCD}_{SB} = 19/9 \pi^2T^3$ is the Stefan-Boltzmann entropy density. Thus, by inverting the $f(g^2)$ function, the coupling constant $g^2$ can be directly obtained from the parametrization of lQCD data for the entropy density $s(T,\mu_B=0)$ at zero baryon chemical potential:
\begin{equation}
\label{eq:g2mub0}
  g^2(T,\mu_B=0) = d \left( (s(T,0)/s^{QCD}_{SB})^e - 1 \right)^f.
\end{equation}
Here $d = 169.934, e = -0.178434$, and $f = 1.14631$ are the dimensionless parameters obtained by adjusting the quasiparticle entropy density $s(T,\mu_B=0)$ to the lQCD data provided by the BMW Collaboration \cite{Borsanyi:2012cr,Borsanyi:2013bia}. 

\begin{figure}[h!]
  \centering
  \includegraphics[width=\columnwidth]{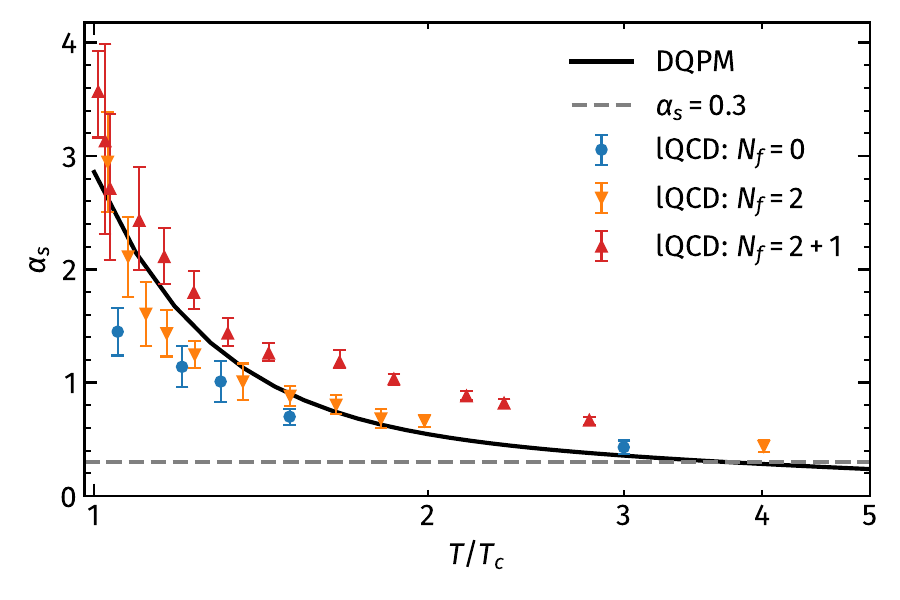}
  \caption{
    Running coupling constant $\alpha_s$ as a function of $T/T_c$ for $\mu_B=0$. The black solid line corresponds to the DQPM running coupling defined by Eq. \eqref{eq:coupling}, and the gray dashed line indicates a constant value of 0.3. The lattice results for quenched QCD, $N_f = 0$ (blue circles), are taken from Ref. \cite{Kaczmarek:2004gv}, for $N_f = 2$ (orange inverted triangles) from Ref. \cite{Kaczmarek:2005ui}, and for $N_f =2 +1$ (red triangles) from Ref. \cite{Kaczmarek:2007pb}.
  }
  \label{fig:alphas}
\end{figure}

Figure \ref{fig:alphas} displays the running coupling constant $\alpha_s$ for $\mu_B=0$ as a function of $T/T_c$, represented by the black solid line in accordance with Eq. \eqref{eq:g2mub0}. The lattice results for quenched QCD, $N_f = 0$ (blue circles) are taken from Ref. \cite{Kaczmarek:2004gv}, for $N_f = 2$ (orange inverted triangles) from Ref. \cite{Kaczmarek:2005ui}, and for $N_f = 2+1$ (red triangles) from Ref. \cite{Kaczmarek:2007pb}. The fit obtained from the entropy density for the DQPM $\alpha_s$ shows a significant increase approaching $T_c$ comparable to the predictions from lQCD. 

It is important to note that the value of the coupling constant obtained by lQCD (i.e., lQCD \textquote{data} in Fig. \ref{fig:alphas}) depends on the definition of $\alpha_s$ extracted from the static potential \cite{Kaczmarek:2004gv, Kaczmarek:2007pb}. Moreover, the DQPM  $\alpha_s$ accounts for nonperturbative effects which make it larger compared to the analytical two- or one-loop running constant \cite{Caswell:1974gg} approaching low temperatures. The gray dashed line in the figure represents a constant value of $\alpha_s =0.3$ frequently used in perturbative QCD models, particularly in the BAMPS calculations.

The extension of the coupling constant to finite baryon chemical potential $\mu_B$ is realized using a scaling hypothesis \cite{Cassing:2008nn}, which works up to $\mu_B \approx 500$ MeV. It assumes that $g^2$ is a function of the ratio of the effective temperature $T^* = \sqrt{T^2 + \mu_q^2/\pi^2}$ and the $\mu_B$-dependent critical temperature $T_c(\mu_B)$ as
\begin{equation}
\label{eq:coupling}
  g^2(T/T_c,\mu_B) = g^2\left(\frac{T^*}{T_c(\mu_B)},\mu_B =0 \right),
\end{equation}
with $T_c(\mu_B) = T_c \sqrt{1 - \alpha\mu_B^2}$, where $T_c$ is the critical temperature at vanishing chemical potential ($\approx$ 0.158~GeV) and $\alpha \simeq 0.974$ GeV$^{-2}$ \cite{Berrehrah:2016vzw}. This extension of the coupling constant for finite $\mu_B$ increases the $\mu_B$ dependence of pole masses and widths, especially in the vicinity of $T_c$ \cite{Soloveva:2020ozg}.

\begin{figure}[ht!]
  \centering
  \includegraphics[width=0.96\columnwidth]{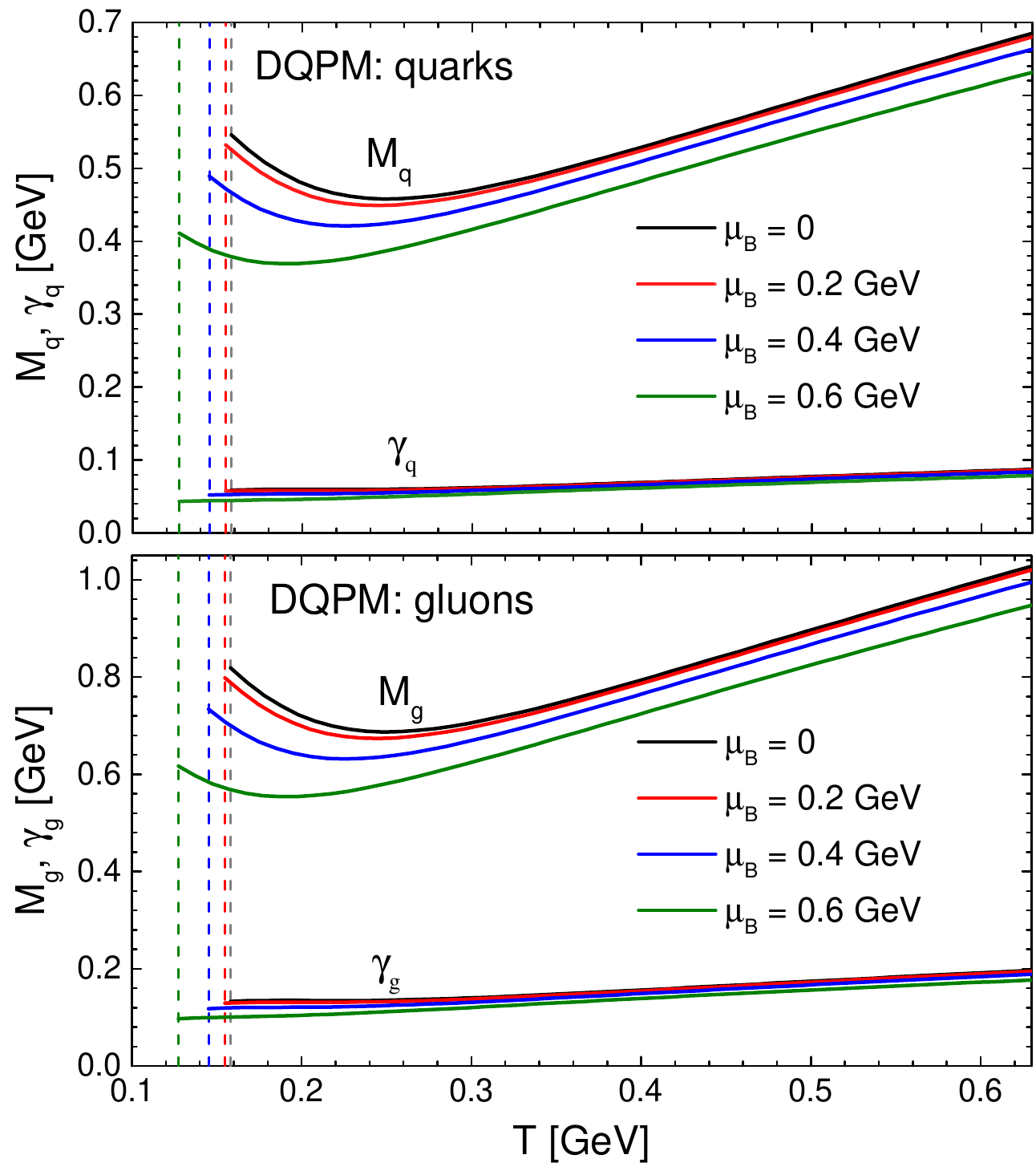}
	\caption{
    The effective quark (upper plot) and gluon (lower plot) masses $M$ and widths $\gamma$ [from the parametrizations in Eqs. (\ref{eq:Mq}), (\ref{eq:Mg}), and (\ref{eq:widths})] as a function of the temperature $T$ for different $\mu_B$. The vertical dashed lines correspond to the DQPM $\mu_B$-dependent critical temperature $T_c(\mu_B)$.
  }
  \label{fig:Masses}
\end{figure}

The actual pole masses of quarks, $M_q$, and gluons, $M_g$, as well as their widths, $\gamma_q$ and $\gamma_g$, are depicted in Fig. \ref{fig:Masses} as a function of $T$ for different baryon chemical potentials $\mu_B=0, 0.2, 0.4$, and 0.6 GeV. As seen from the figure, the masses grow with $T$ and decrease with increasing $\mu_B$ and move to the low $T$ following the $\mu_B$-dependent critical temperature $T_c(\mu_B)$ (depicted as dashed vertical lines). The widths of quarks and gluons increase with $T$ and show only a weak $\mu_B$ dependence. For $\mu_B =0$ the DQPM gives $M_q = \frac{2}{3} M_g, \gamma_q = \frac{4}{9} \gamma_g$.

Having the quasiparticle properties, dressed propagators, and coupling constant as given by the DQPM, one can evaluate the scattering amplitudes as well as the cross sections and the transport coefficients of quarks and gluons in the QGP as a function of the temperature and the chemical potential \cite{Berrehrah:2013mua,Moreau:2019vhw}.


\subsection{Matrix elements in the DQPM}
\label{sec:DQPM_ME}

In order to calculate the matrix elements corresponding to a scattering of DQPM partons, the scalar propagator has to be replaced by the following propagators -- with full Lorentz structure -- to describe a massive vector gluon and massive (spin $1/2$) fermion with a finite width \cite{Berrehrah:2013mua}:
\begin{equation}
\label{eq:prop_g}
  \includegraphics[width=0.2\columnwidth]{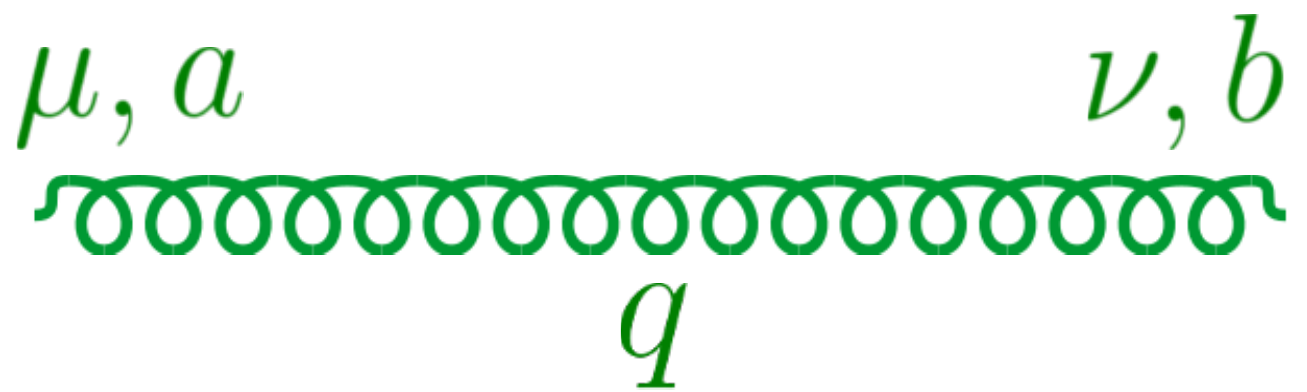} = -i \delta_{ab} \frac{g^{\mu \nu} - q^\mu q^\nu / M^2_g}{q^2 - M^2_g + 2i \gamma_g q_0},
\end{equation}

\begin{equation}
\label{eq:prop_q}
  \includegraphics[width=0.2\columnwidth]{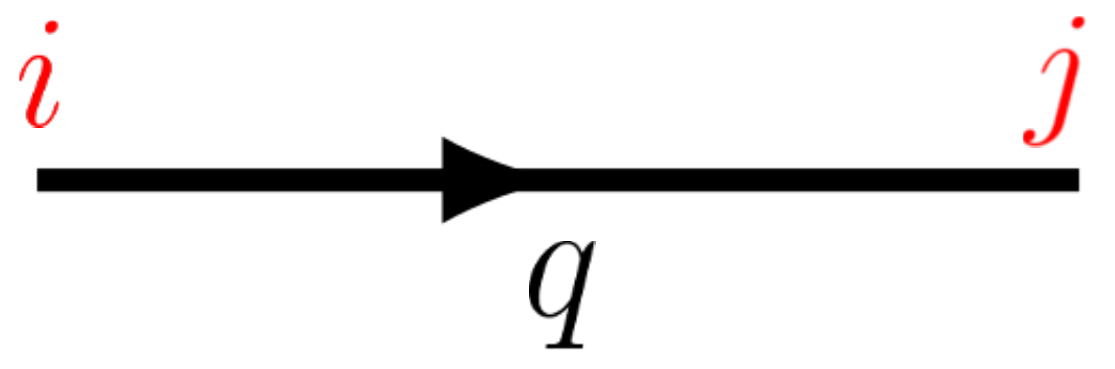} = i \delta_{ij} \frac{\slashed{q} + M_q}{q^2 - M^2_q + 2i \gamma_q q_0},
\end{equation}
where $q$ is the 4-momentum of the exchanged particle. The $\delta$ functions ensure that the exchanged quark or gluon is connected with other parts of the diagram with the same color ($a,b$ for the gluon and $i,j$ for the quark); i.e., color and spin during the propagation is conserved.

In the DQPM, only transverse gluons are included in the calculation of thermodynamic quantities, in that the contribution of hard longitudinal gluons is found to be negligible \cite{Blaizot:2000fc,Peshier:2000hx}. The details on the summation over gluon polarization for massive gluons are given in Appendix \ref{appx:gluon_polarization}. We note that we use the covariant gauge for the evaluation of cross sections.

For the calculation of elastic quark-quark and quark-gluon scatterings we refer the reader to Refs. \cite{Moreau:2019vhw,Soloveva:2020hpr,Grishmanovskii:2022tpb}.


\section{Gluon radiation processes in the DQPM}

Now we step to the calculation of the inelastic processes for gluon radiation from thermal quark-quark ($q+q$) and quark-gluon ($q+g$) scatterings: $q+q \to q+q +g$ and $q+g \to q+g+g$ within leading-order Feynman diagrams with effective propagators and vertices from the DQPM as specified in Sec. \ref{sec:DQPM_ME}. We note that here and further on when noting reactions with \textquote{quark} we also imply the corresponding reactions for antiquarks (if not specified explicitly).


\subsection{Matrix elements for \texorpdfstring{$2\to 3$}{2 -> 3} reactions}

The inelastic scattering is accounted for by the leading order Feynman diagrams for $2\to 3$ processes considering $t$, $u$, and $s$ channels. All diagrams are calculated explicitly without assumptions, accounting for all possible interferences between different channels.
 
All processes are computed for the \textquote{on-shell} case, when the masses of final partons, as well as the radiative gluon, are taken at the pole position of their spectral functions since \textquote{off-shell} calculations including the integration over their spectral functions are much more computationally expensive. We note that on-shell elastic cross sections differ from off-shell ones by less than 20\% \cite{Grishmanovskii:2022tpb}.


\subsubsection{Gluon radiation from quark + quark scattering}
\label{sec:ampl_qq}

\begin{figure}[t!]
  \centering
  \includegraphics[width=\columnwidth]{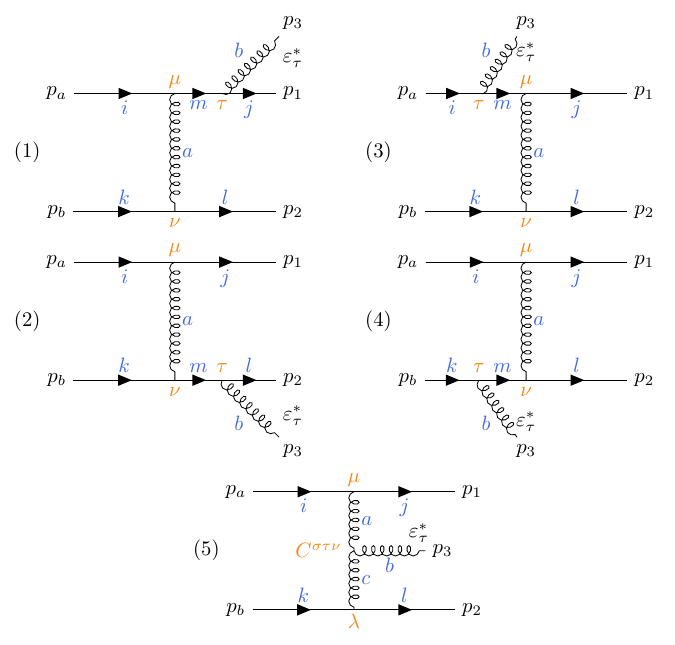}
  \caption{Feynman diagrams for the $t$ channel of the $q+q \to q+q +g$ scattering.}
\label{fig:diags_qq}
\end{figure}

We start with gluon radiation from quark + quark scattering, $q+q \to q+q +g$. The corresponding Feynman diagrams for $t$ channels are presented in Fig. \ref{fig:diags_qq}. Here $p_a, p_b$ and $p_1, p_2$ stand for the 4-momenta of initial and final quarks, respectively, while $p_3$ corresponds to the 4-momentum of the emitted gluon.

Denoting
\begin{align}
  &\GP{\mu}{\nu}{p} = \left[-i\frac{g_{\mu\nu}-(p_\mu p_\nu) / M_g^2}{p^2-M_g^2+2 i \gamma_g \omega_p}\right] \quad \text{(gluon propagator)},
  \nonumber\\
  &\QP{p} = \left[i \frac{\slashed{p}+M_q}{p^2-M_q^2 + 2i\gamma_q \omega_{p}}\right]  \quad \text{(quark propagator)},
  \nonumber\\
  &\VX{\nu}{a}{ik} = \left(-ig\gamma^{\nu} T^a_{ik}\right) \quad \text{(quark-gluon vertex)},
  \nonumber\\
  &\GV{abc}{\mu\nu\tau}{p_1,p_2,p_3} = -g f^{abc} C^{\mu\nu\tau}(p_1,p_2,p_3)\nonumber\\ &~~~~~~~~~~~~~~~~~~~~~~~~~~~~~~~~\text{(three gluon vertex)},
\end{align}
where $\mu,\nu$ stand for the Lorentz indices, $i,k = 1,\dots,3$ for the quark color indices and $a,c,b = 1,\dots,8$ for the gluon color indices, one can write the invariant amplitudes $i \mathcal{M}_i = i \mathcal{M}_i (q^i q^k \to q^j q^l g^b)$ -- corresponding to the Feynman diagrams in Fig. \ref{fig:diags_qq} -- as
\begin{widetext}
\begin{align}
  i \mathcal{M}_1 &=
  \ubar^l(p_2) \VX{\nu}{a}{lk} u^k(p_b) \GP{\mu}{\nu}{p_b-p_2}
  \ubar^j(p_1) \varepsilon_{\tau}^{*}(p_3) \VX{\tau}{b}{jm}
  \QP{p_1+p_3}
  \VX{\mu}{a}{mi} u^i(p_a),
  \label{eq:M}\\
  i \mathcal{M}_2 &=
  \ubar^j(p_1) \VX{\mu}{a}{ji} u^i(p_a) \GP{\mu}{\nu}{p_a-p_1}
  \ubar^l(p_2) \varepsilon_{\tau}^{*}(p_3) \VX{\tau}{b}{lm}
  \QP{p_2+p_3}
  \VX{\nu}{a}{mk} u^k(p_b),
  \nonumber\\
  i \mathcal{M}_3 &=
  \ubar^l(p_2) \VX{\nu}{a}{lk} u^k(p_b) \GP{\mu}{\nu}{p_b-p_2}
  \ubar^j(p_1) \VX{\mu}{a}{jm}
  \QP{p_a-p_3}
  \varepsilon_{\tau}^{*}(p_3) \VX{\tau}{b}{mi} u^i(p_a),
  \nonumber\\
  i \mathcal{M}_4 &=
  \ubar^j(p_1) \VX{\mu}{a}{ji} u^i(p_a) \GP{\mu}{\nu}{p_a-p_1}
  \ubar^l(p_2) \VX{\nu}{a}{lm} 
  \QP{p_b-p_3}
  \varepsilon_{\tau}^{*}(p_3) \VX{\tau}{b}{mk} u^k(p_b),
  \nonumber\\
  i \mathcal{M}_5 &=
  \ubar^j(p_1) \VX{\mu}{a}{ji} u^i(p_a) \; \ubar^l(p_2) \VX{\lambda}{c}{lk} u^k(p_b)
  \GP{\mu}{\nu}{p_a-p_1}
  \GP{\lambda}{\sigma}{p_b-p_2} \GL{*}{\tau}{p_3}
  \GV{abc}{\sigma\tau\nu}{p_b-p_2,-p_3,p_2-p_b+p_3}. \nonumber
\end{align}
\end{widetext}

The invariant amplitude squared -- averaged over initial states of spin and color and summed over final states -- is
\begin{align}
  \Msq{q+q\prime \to q+q\prime+g} = \frac{1}{N_c^2} \sum_{\text{color}}\frac{1}{(2s_q+1)(2s_{q\prime}+1)}
  \nonumber\\ \times
  \sum_{\text{spin}}
  |\mathcal{M}_1+\mathcal{M}_2+\mathcal{M}_3+\mathcal{M}_4+\mathcal{M}_5|^2,
  \label{eq:Mbar}
\end{align}
where $q^\prime$ denotes a quark with a possible different flavor than the $q$ quark.

There are $5 \times 5=25$ possible combinations of multiplications, which, in turn, can be reduced to 15 using symmetry relations: $|\bar M_{ij}| = |\bar M_{ji}|$, where $ |\bar M_{ij}| \sim M_i^* M_j$ (cf. Eq. (\ref{eq:Mbar})). An explicit expression for the squared amplitude cannot be written in a compact form and therefore the analysis of this expression can only be performed numerically. In this study the final analytical expression was evaluated using the FeynCalc package \cite{Mertig:1990an,Shtabovenko:2016sxi}.

In general, the $q+q$ reaction implies different channels depending on the flavors of the interacting quarks. The only channel that is common for all $q+q$ reactions is the $t$ channel. \\
The $u$ channel is relevant only for quarks with the same flavor, and even though the angular distribution for this reaction is different, the integrated cross section is of the same value as for the quarks with different flavors. \\
The $s$ channel is relevant for the $q + \bar{q}$ reaction and for the $2 \to 3$ reaction is expected to be strongly suppressed for any value of $\srs$. \\
Therefore, in this study, we consider only the $t$ channel for $q+q$ reactions, noting that the corrections given by the other channels are negligible.


\subsubsection{Gluon radiation from quark + gluon scattering}

\begin{figure*}[th!]
    \centering
    \includegraphics[width=0.9\textwidth ]{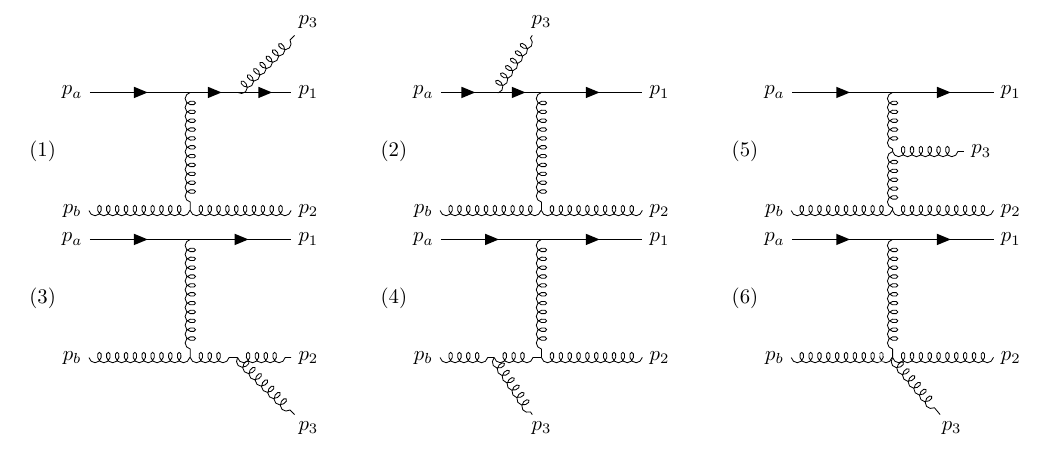}
    \includegraphics[width=0.49\textwidth]{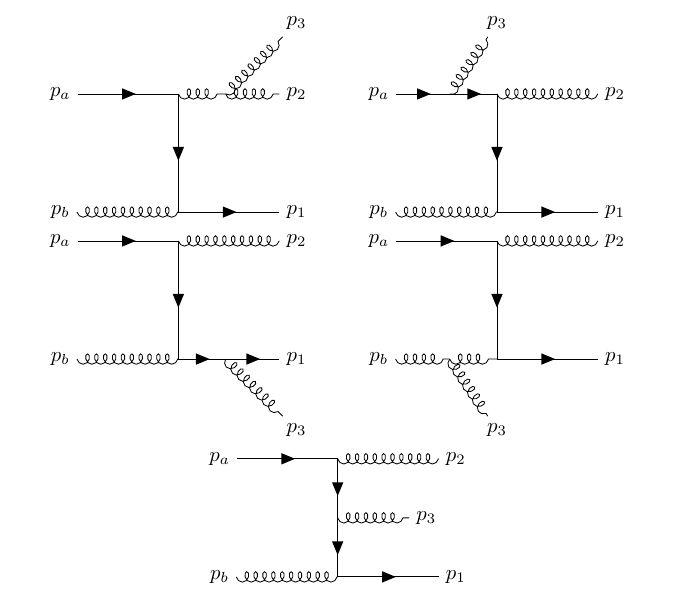} 
    \includegraphics[width=0.49\textwidth]{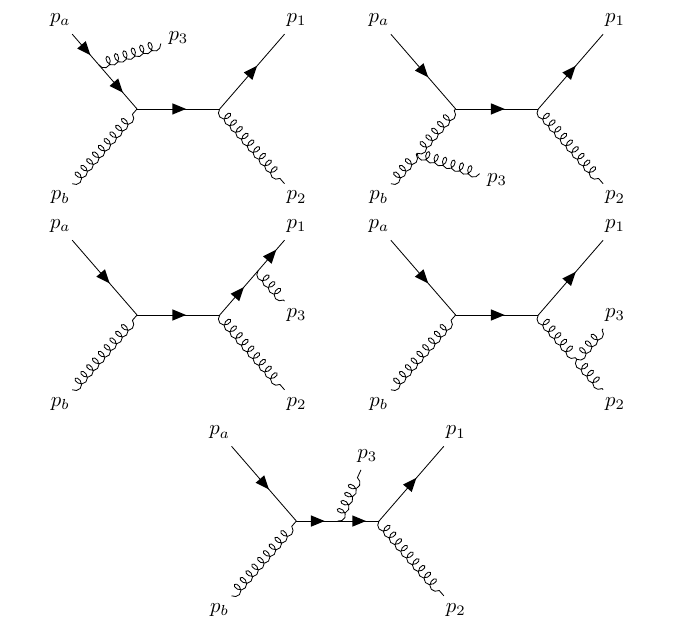}
    \caption{
      Feynman diagrams for $t$ channels (upper six diagrams), $u$ channels (lower left five diagrams), and $s$ channels (lower right five diagrams) for the $q+g \to q+g+ g$ process.
    }
\label{fig:diags_gq}
\end{figure*}

We continue with gluon radiation from quark + gluon scattering $q+g \to q+g +g$. Contrary to $q+q$ reactions, $q+g$ reactions incorporate all $t$, $u$, and $s$ channels regardless of the quark flavor. However, we will show explicitly that, as in the case of the $q+q$ reaction, the $t$ channel plays the dominant role in the $q+g$ reactions, while contributions from the other channels could as well be omitted. The corresponding Feynman diagrams for $t$, $u$, $s$ channels are presented in Fig. \ref{fig:diags_gq}.

For the $t$ channel of the $q+g$ reaction the matrix amplitudes (according to the numeration in the upper two rows in Fig. \ref{fig:diags_gq}) read as
\begin{widetext}
\begin{align}
  i \mathcal{M}_1 &= 
  \ubar(p_1) 
  \VX{\mu}{d}{kn} 
  \QP{p_1+p_3, m_q} 
  \VX{\lambda}{f}{ni} 
  u(p_a) 
  \GV{bef}{\sigma\tau\nu}{p_b-p_2,-p_b,p_2}
  \GP{\lambda}{\sigma}{p_2-p_b} \GL{*}{\mu}{p_3} \GL{*}{\nu}{p_2} \GL{}{\tau}{p_b},
  \label{eq:Mglue}\\
  i \mathcal{M}_2 &= 
  \ubar(p_1) 
  \VX{\sigma}{f}{kn} 
  \QP{p_1+p_2-p_b, m_q} 
  \VX{\mu}{d}{ni} 
  u(p_a) 
  \GV{bef}{\lambda\tau\nu}{p_b-p_2,-p_b,p_2}
  \GP{\lambda}{\sigma}{p_2-p_b} \GL{*}{\mu}{p_3} \GL{*}{\nu}{p_2} \GL{}{\tau}{p_b},
  \nonumber\\
  i \mathcal{M}_3 &= 
  \ubar(p_1) 
  \VX{\lambda}{f}{ki} 
  u(p_a) 
  \GV{bfn}{\nu\mu\xi}{-p_2,-p_3,p_2+p_3}
  \GV{deh}{\sigma\tau\rho}{-p_b+p_2+p_3,p_b,-p_2-p_3}
  \nonumber\\ &\times
  \GP{\rho}{\xi}{p_2-p_b} 
  \GP{\lambda}{\sigma}{p_2-p_b} \GL{*}{\mu}{p_3} \GL{*}{\nu}{p_2} \GL{}{\tau}{p_b},
  \nonumber\\
  i \mathcal{M}_4 &= 
  \ubar(p_1) 
  \VX{\lambda}{f}{ki} 
  u(p_a) 
  \GV{efh}{\tau\mu\rho}{-p_b,p_3,p_b-p_3}
  \GV{bdh}{\nu\sigma\xi}{p_2,p_b-p_2-p_3,-p_b+p_3}
  \nonumber\\ &\times
  \GP{\rho}{\xi}{p_2-p_b} 
  \GP{\lambda}{\sigma}{p_2-p_b} \GL{*}{\mu}{p_3} \GL{*}{\nu}{p_2} \GL{}{\tau}{p_b},
  \nonumber\\
  i \mathcal{M}_5 &= 
  \ubar(p_1) 
  \VX{\lambda}{f}{ki} 
  u(p_a) 
  \GV{dfh}{\sigma\mu\xi}{-p_b+p_2+p_3,-p_3,-p_2+p_b}
  \GV{beh}{\nu\tau\rho}{-p_2,p_b,-p_b+p_2}
  \nonumber\\ &\times
  \GP{\rho}{\xi}{p_2-p_b} 
  \GP{\lambda}{\sigma}{p_2-p_b} \GL{*}{\mu}{p_3} \GL{*}{\nu}{p_2} \GL{}{\tau}{p_b},
  \nonumber\\
  i \mathcal{M}_6 &= 
  \ubar(p_1) 
  \VX{\lambda}{f}{ki} 
  u(p_a) 
  \Gamma^{fbde}_{\mu\nu\sigma\tau}
  \GP{\lambda}{\sigma}{p_2-p_b} \GL{*}{\mu}{p_3} \GL{*}{\nu}{p_2} \GL{}{\tau}{p_b}. 
  \nonumber
\end{align}
\end{widetext}

We omit showing explicitly the expressions for $u$ and $s$ channels, noting that they can be expressed in a similar way.


\subsection{Cross section for \texorpdfstring{$2 \to 3$}{2 -> 3} reactions}

The cross section for a $2\to 3$ process $(a+b \to 1+2+3)$ is given by
\begin{align}
  \sigma_{2\to 3} &= \frac{1}{F} 
  \int \frac{d^3 p_1}{\dpi^3 2E_1} 
  \int \frac{d^3 p_2}{\dpi^3 2E_2} 
  \int \frac{d^3 p_3}{\dpi^3 2E_3} 
  \nonumber\\ &\times
  \dpi^4 \delta^{(4)}(p_a + p_b - p_1 - p_2 - p_3)
  \Msq{2\to3},
\end{align}
where the flux is $F = 4p_C\srs$ with $p_C$ denoting the initial 3-momentum in the center-of-mass (c.m.) frame.

This integration implies, in general, a nine-dimensional integration. However, it is numerically more efficient to analytically integrate out the $\delta$ function to end up with a five-dimensional integration. The further choice of independent phase space variables depends on the observables we are interested in. The first possible choice to perform the integration is to consider the $2 \to 3$ reaction as a sum $(2 \to 2) + (1 \to 2)$ of processes by introducing the intermediate \textquote{state} $p_x = p_2 + p_3$ (or $p_x = p_1 + p_2$). In this case, the set of independent variables can be chosen as $\{m_x,\theta,\varphi,\xi,\psi\}$, where $\{\theta,\varphi\}$ are polar and azimuthal angles between $\vec{p}_1$ and $\vec{p}_x$ (or $\vec{p}_3$ and $\vec{p}_x$) in the c.m. frame, $\{\xi,\psi\}$ are polar and azimuthal angles of $\vec{p}_2$ and $\vec{p_3}$ (or $\vec{p_1}$ and $\vec{p_2}$) in the $p_x$ rest frame and $m_x = \sqrt{p_x^2}$. The integration over $\varphi$ can be done analytically due to the symmetry of the system giving the final expression for the total cross section (see Appendix \ref{appx:var1} for details):
\begin{align}
  \label{eq:XS_2to3_var1}
  \sigma_{2\to3} &= \frac{1}{8 \dpi^5 \sqrt{s} F} \int_0^1 d\cos\theta \int_0^1 d\cos\xi \int_0^{2\pi}d\psi
  \nonumber \\ &\times
  \int_{m_2+m_3 (m_1+m_2)}^{\sqrt{s}-m_1 (\srs-m_3)} dm_x \ p_{x,C} \ p_{2,X} \Msq{2\to3}.
\end{align}
This choice of variables allows to investigate the angular distribution of the inelastic cross section, in particular $d\sigma/d\cos\theta$ with $\theta$ being the scattering angle of $p_1$ (leading quark) or $p_3$ (emitted gluon) in the c.m. frame. Also this set of variables is convenient when one wants to implement the cross sections into transport simulations.

The second choice of variables (in the c.m. frame) implies switching to \textquote{GB coordinates}, which were used by Gunion and Bertsch in Ref. \cite{Gunion:1981qs} and in the BAMPS study \cite{Fochler:2013epa}, $(q_t,k_t,y,\varphi)$, where $q_t$ is the transverse momentum transfer, $k_t$ is the transverse momentum of the emitted gluon, $y$ is the rapidity of the emitted gluon, and $\varphi$ is the angle between $k_t$ and $q_t$. In these variables the $2 \to 3$ cross section can be expressed as
\begin{align}
  \label{eq:XS_2to3_var2}
  \sigma_{2\to 3} &= \frac{1}{128 \pi^4 F} \int_{0}^{p_C^2} dq_t^2 \int_0^{p_C^2} dk_t^2 \int_{y_{\min}}^{y_{\max}} dy \int_0^\pi d\varphi 
  \nonumber \\ &\times
  \Msq{2\to 3} \sum\left( \left.\frac{\partial \mathcal{F}}{\partial y_1}\right|_{\mathcal{F}=0} \right)^{-1},
\end{align}
where $y_1$ denotes the rapidity of particle 1 and $\varphi$ is the angle between $\vec{q}_t$ and $\vec{k}_t$. The available phase space limits the rapidity to $y_{\max/\min}=\pm \text{arccosh}(\frac{\sqrt{p_C^2+m_3^2}}{m_{3,t}})$. The function $\mathcal{F}$ comes from the transformation of the $\delta$ function (see Appendix \ref{appx:var2}).

We have numerically checked that both Eqs. \eqref{eq:XS_2to3_var1} and \eqref{eq:XS_2to3_var2} give the same results.


\section{Results}
\label{sec:results}

We proceed with showing the results from numerical calculations of cross sections for radiative processes $q+q\to q+q+g$ and $q+g\to q+g+g$. We note that all calculations here and below are done for $\mu_B=0$. If not stated otherwise, all parton masses and widths are chosen to be thermal DQPM masses and widths \cite{Moreau:2019vhw}, which depend on temperature and $\mu_B$. The parton masses are taken at the pole position of the spectral functions (cf. Fig. \ref{fig:Masses}), i.e.,
\begin{align}
  m_a &= m_b = m_1 = m_2 = M_q = M^{DQPM}_q(T,\mu_B),\nonumber\\
  m_3 &= M_g = M^{DQPM}_g(T,\mu_B),\nonumber\\
  \gamma_q &= \gamma^{DQPM}_q(T,\mu_B), \nonumber\\
  \gamma_g &= \gamma^{DQPM}_g(T,\mu_B).
\end{align}
We mention that when indicating the processes $q+q\to q+q$ and $q+q\to q+q+g$ we imply the reactions of light quarks of different flavors, too. The same convention holds for the scatterings including antiquarks.


\subsection{pQCD limit of the DQPM}

We start with showing explicitly that the DQPM amplitudes reproduce the pQCD amplitudes \cite{Berends:1981rb,Ellis:1985er} in the \textquote{pQCD} limit of massless partons with zero widths.

\begin{figure}[ht!]
  \centering
  \includegraphics[width=\columnwidth]{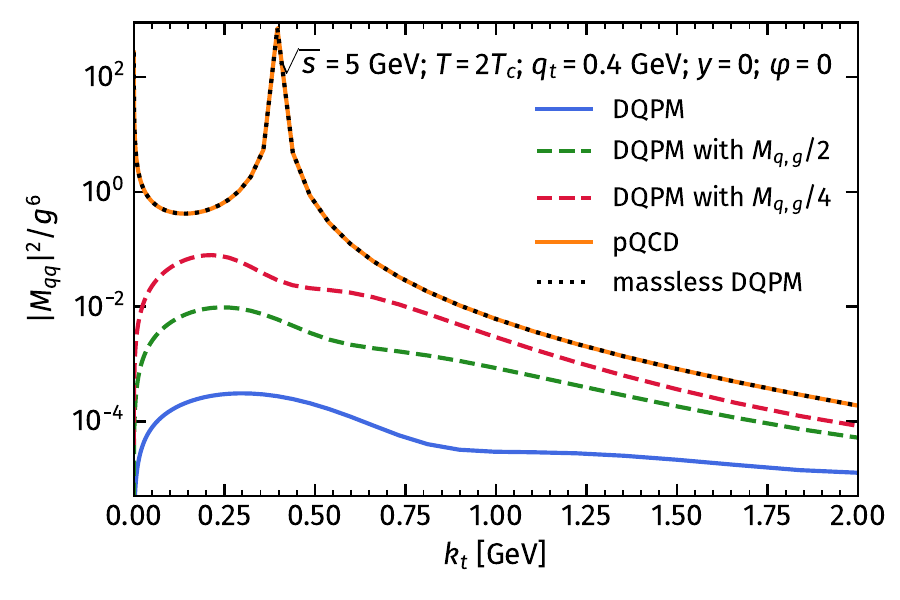}
  \caption{
    Squared scattering amplitude (scaled by $g^6$) for the $q+q \to q+q+g$ process for $\sqrt{s}=5$ GeV as a function of the transverse momentum of the emitted gluon $k_t$ in the case of the full DQPM (calculated with pole masses $M_q, M_g$) for $T=2T_c$ (blue line), pQCD result from Ref. \cite{Berends:1981rb} (orange line), and DQPM in the massless limit (black dotted line). The dashed green and red lines display the DQPM amplitudes calculated with reduced pole masses of quarks and gluons $M_q/n$ and $M_g/n, \ n=2, 4$.
  }
  \label{fig:d4XS-kt}
\end{figure}

Figure \ref{fig:d4XS-kt} shows the squared amplitude for the inelastic squared amplitude (scaled by $g^6$) for the $q+q \to q+q+g$ process as a function of the transverse momentum of the emitted gluon $k_t$ in the case of the full DQPM (blue line), pQCD result from Ref. \cite{Berends:1981rb} (orange line), and DQPM in the massless limit (black dotted line). The quantity $|M_{qq}|^2/g^6$ is computed for $q_t=0.4$ GeV, $y=0$, and $\varphi=0$.

The convergence of the \textquote{massless DQPM} and \textquote{pQCD} results illustrates the correspondence between the pQCD model and the DQPM in the massless limit as it follows from the form of the quark [Eq. \eqref{eq:prop_q}] and gluon [Eq. \eqref{eq:prop_g}] propagators. One can also see that the pQCD amplitude diverges in the infrared ($k_t \to 0$) and collinear ($\vec{k}_t = \vec{q}_t$) limits, which makes it impossible to calculate the integrated cross section without introducing the corresponding cutoffs. The result from the full DQPM model calculated with the pole masses of quarks and gluons (blue line) shows the absence of these divergencies and therefore allows for the calculation of the integrated cross section. The dashed green and red lines display the DQPM amplitudes calculated with reduced masses of quarks and gluons $M_q/n$ and $M_g/n, \ n=2, 4$. One can see that the reduction of the DQPM masses leads to an enhancement of the DQPM amplitudes that approach the pQCD limit.


\subsection{Angular distributions}

\begin{figure}[ht!]
  \centering
  \includegraphics[width=\columnwidth]{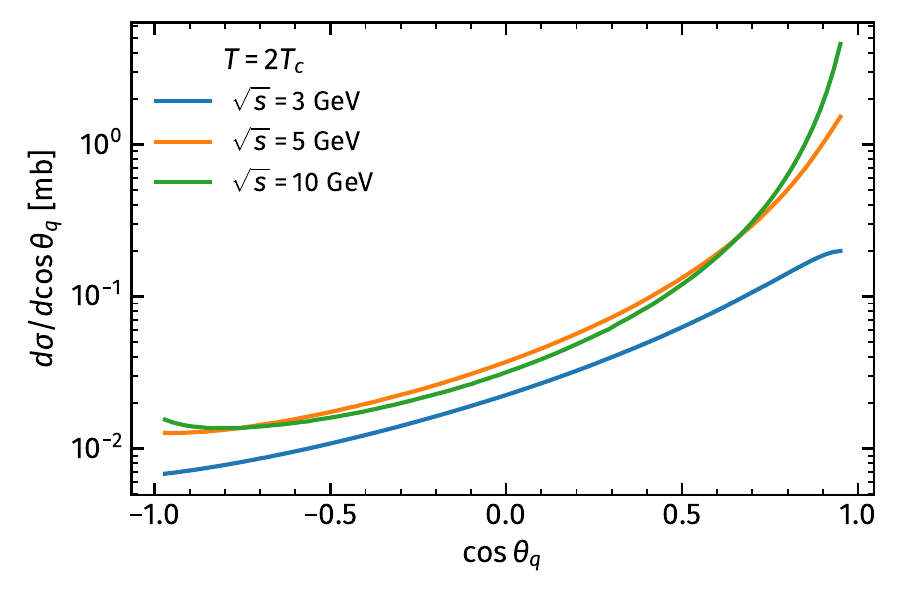}
  \caption{
    Inelastic differential cross sections for $q+q \to q+q+g$ scattering for light quarks of different flavor for $T = 2T_c$ in the c.m. frame as a function of the angle $\theta_q$ between the initial and final momentum of the quark. The different lines correspond to different collision energies $\srs$ from 3 to 10 GeV (see legend).
  }
  \label{fig:dXSi-cosTh_q}
\end{figure}

\begin{figure}[ht!]
  \centering
  \includegraphics[width=\columnwidth]{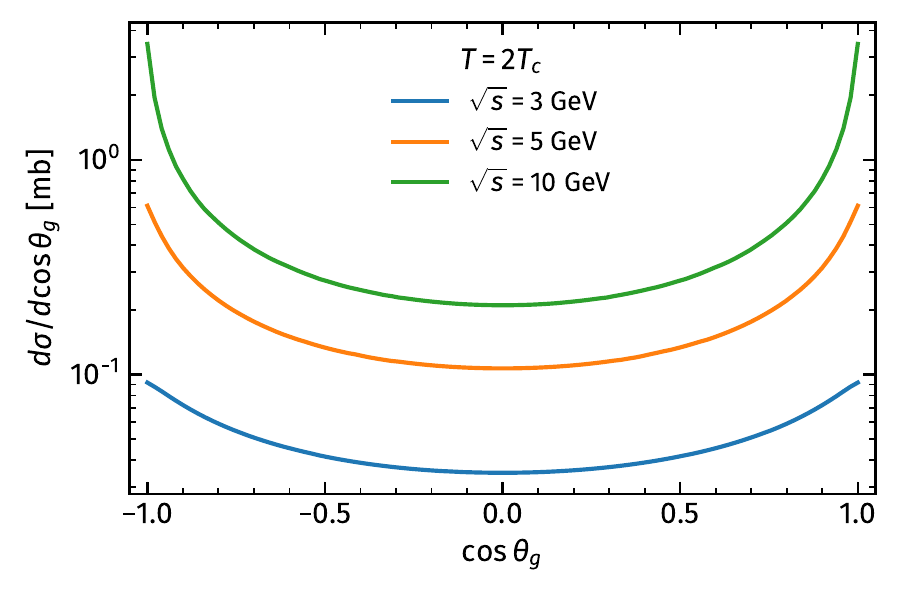}
  \caption{
    Inelastic differential cross sections for $q+q \to q+q+g$ processes for $T = 2T_c$ as a function of the angle $\theta_g$ between the momentum of the emitted gluon relative to the collision axis in the c.m. frame. The different lines correspond to different collision energies $\srs$ from 3 to 10 (see legend).
  }
  \label{fig:dXSi-cosTh_g}
\end{figure}

Figure \ref{fig:dXSi-cosTh_q} shows the inelastic differential cross sections for $q+q \to q+q+g$ scattering for light quarks of a different flavor for $T = 2T_c$ as a function of the angle $\theta_q$ between the outgoing quark after emission of the gluon relative to the collision axis in the c.m. frame for different collision energies $\srs$. As one can see the angular distribution becomes steeper with increasing collision energy; thus for high-energy scatterings a forward scattering to small angles dominates.

Figure \ref{fig:dXSi-cosTh_g} shows the inelastic differential cross sections for $q+q \to q+q+g$ processes for $T = 2T_c$ as a function of the angle $\theta_g$ between the momentum of the emitted gluon relative to the collision axis in the c.m. frame for different collision energies $\srs$. All lines are symmetric with respect to the angle $\theta_g$ meaning that forward and backward directions for the emitted gluon are equal. At the same time, the distribution shows rapid growth of the cross section for small $\theta_g$ with increasing energy $\srs$ indicating that small angles are dominant for the gluons as well as for the quarks.


\subsection{Transverse momentum distributions, comparison of DQPM and BAMPS}

\begin{figure}[ht!]
  \centering
  \includegraphics[width=\columnwidth]{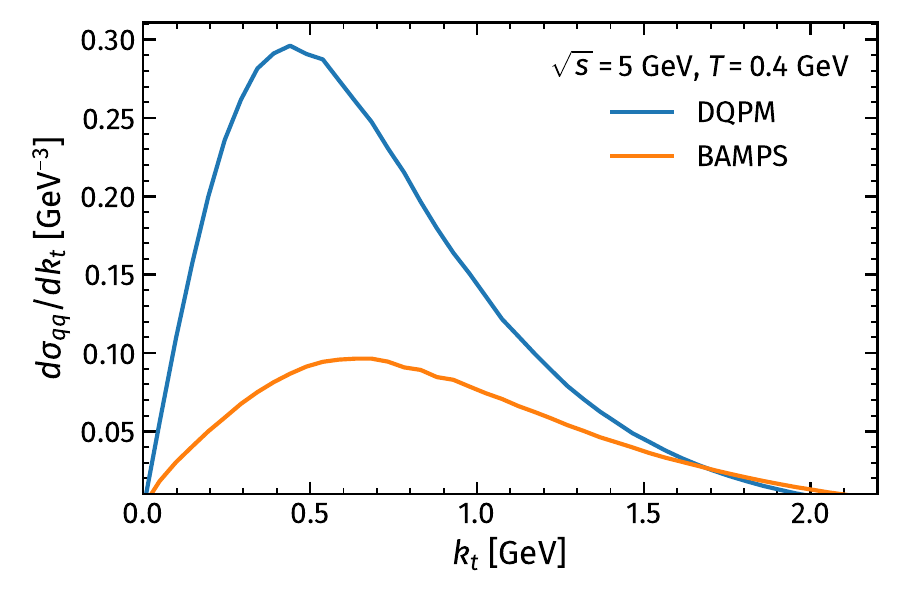}
  \includegraphics[width=\columnwidth]{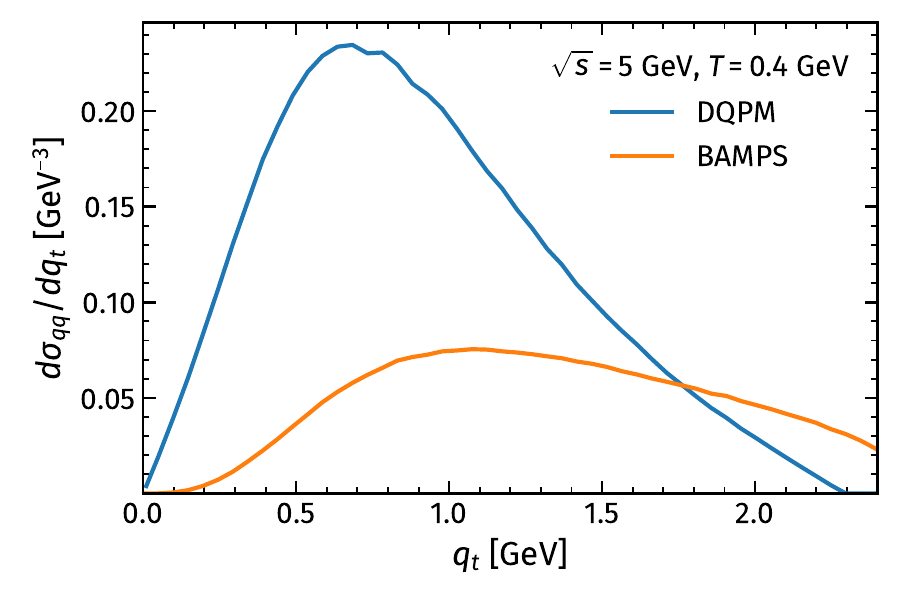}
  \caption{
    Differential cross section $d\sigma/dk_t$ as a function of $k_t$ (upper plot) and $d\sigma/dq_t$ as a function of $q_t$ (lower plot) for the $q+q \to q+q+g$ scattering of light quarks of different flavors calculated within the DQPM (blue lines) and BAMPS model (orange lines) \cite{Fochler:2013epa}.
  }
  \label{fig:dXS-ktqt}
\end{figure}

Figure \ref{fig:dXS-ktqt} shows the differential cross section $d\sigma/dk_t$ as a function of the transverse momentum of the emitted gluon $k_t$ (upper plot) and $d\sigma/dq_t$ (lower plot) versus transverse momentum transfer $q_t$ (lower plot), respectively, for the $q+q \to q+q+g$ process for light quarks of different flavors calculated with the DQPM (solid blue lines) matrix elements.

We compare the DQPM transverse momentum distributions in $k_t$ and $q_t$ to the distributions calculated based on the pQCD BAMPS model (orange lines) for the improved Gunion-Bertsch description following Ref. \cite{Fochler:2013epa}. We recall that in the BAMPS model \cite{Fochler:2013epa} the strong-coupling constant is fixed as $\alpha_s = 0.3$; the exchange parton has a Debye mass in the case of a gluon, $m_D^2= \dfrac{8\alpha_s}{\pi}(N_c+N_f)T^2$, while the scattered partons are assumed to be massless. The squared matrix element \cite{Fochler:2013epa} is taken as 
\begin{multline}
\label{eq:M2BAMPS}
	{\left|\overline{\mathcal{M}_{qq \to qqg}}\right|}^2 =
	 \frac{32}{3} g^6	 \frac{s^2}{(q_t^2 + m_D^2)^2} 
	(1-\bar{x})^2 \\ 
	\times 
	\left[ \frac{ {\bf k}_t}{k_t^2 + m_D^2} + \frac{ {\bf q}_t - {\bf k}_t}{({\bf q}_t - {\bf k}_t)^2 + m_D^2} \right]^2 ,
\end{multline}
where ${\bf q}_t, {\bf k}_t$ are the 2-vectors of the transverse momentum of the quark and emitted gluon, respectively. Here the quantity $x$ characterizes the fraction of the light cone momentum carried away by the radiated gluon. It can be related to the rapidity of the emitted gluon by $x = \frac{k_t}{\srs} {\rm e}^y$. In Eq. (\ref{eq:M2BAMPS}) the $k_t \to 0$ divergence (as well as $q_t \to 0$ divergence) of the pQCD amplitude is regularized by an extra Debye mass -- also for the emitted gluon -- which leads to the final cross section.

As seen from Fig. \ref{fig:dXS-ktqt} the DQPM distributions are enhanced at low $k_t$ and $q_t$ momenta compared to the BAMPS results which show broader distributions. Since we selected for the comparison a large $T=0.4$ GeV where the DQPM $\alpha_s(T) \approx 0.3$ (cf. Fig. \ref{fig:alphas}), i.e., similar to $\alpha_s = 0.3$ in BAMPS, the main differences between the DQPM and BAMPS results can be related to the thermal masses in the DQPM versus the zero masses of partons in pQCD and the different propagators. 


\subsection{Channel decomposition for the total cross sections}

We move on to present the DQPM results for the energy and temperature dependence of the total inelastic cross sections. However, first we investigate which channels and which diagrams play a dominant role for the gluon radiation from $q+q$ and $q+g$ scattering.

\begin{figure}[t!]
  \centering
  \includegraphics[width=\columnwidth]{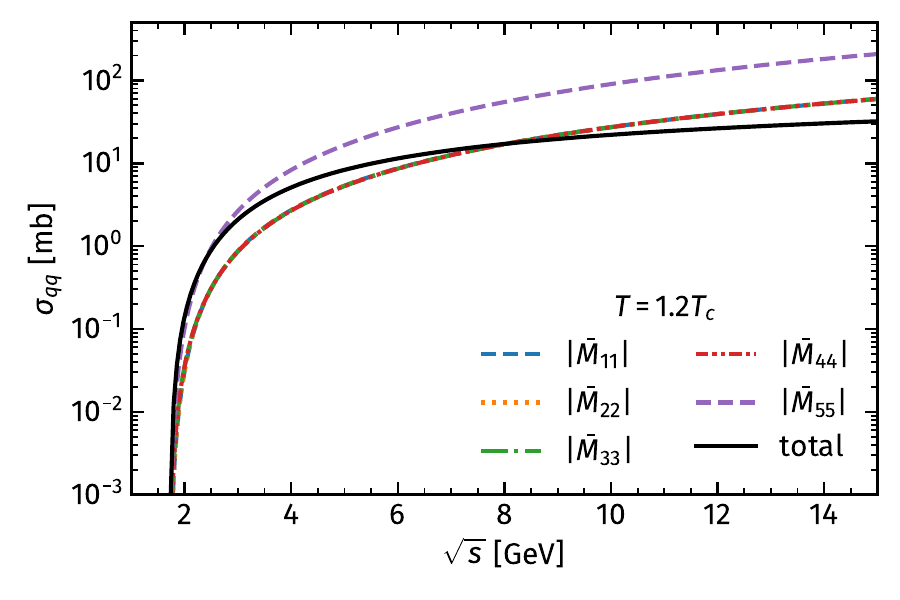}
  \caption{
    Contributions of the squared amplitudes $|\bar M_{ii}|$ (i.e., noninterference terms, shown by the dashed lines of different colors; see the legend) to the total $q+q\to q+q+g$ cross section with the inclusion of all interference terms (solid black line) as a function of the collision energy $\sqrt{s}$ at $T=1.2T_c$.
  }
  \label{fig:XSi_decomp}
\end{figure}

In Fig. \ref{fig:XSi_decomp} we show the contributions of noninterference terms corresponding to the squared amplitude $|\bar M_{ii}|$ [where the amplitudes $M_i$ are defined by Eq. (\ref{eq:M}) and shown by the diagrams 1-5 in Fig. \ref{fig:diags_qq}] (dashed lines of different colors) to the total $q+q\to q+q+g$ cross section (solid black line) calculated including all interference terms as a function of the collision energy $\sqrt{s}$ calculated at $T=1.2T_c$. Here we used the full gluon polarization (cf. Appendix \ref{appx:gluon_polarization}).

One can see that the dominant contribution comes from the term $|\bar M_{55}|$ corresponding to the squared amplitude of diagram 5 in Fig. \ref{fig:diags_qq} with the emission of a gluon from the gluon exchange line. It is larger than the contribution of each of the other diagrams 1-4 at all $\srs$. Diagrams 1, 2, 3, and 4 contribute equally; i.e., $|\bar M_{11}| = |\bar M_{22}| = |\bar M_{33}| = |\bar M_{44}|$. As follows from Fig. \ref{fig:XSi_decomp} the total cross section (solid black line) -- calculated including all terms and their interference -- is smaller than the $|\bar M_{55}|$ term due to destructive interference. 

\begin{figure}[ht!]
  \centering
  \includegraphics[width=\columnwidth]{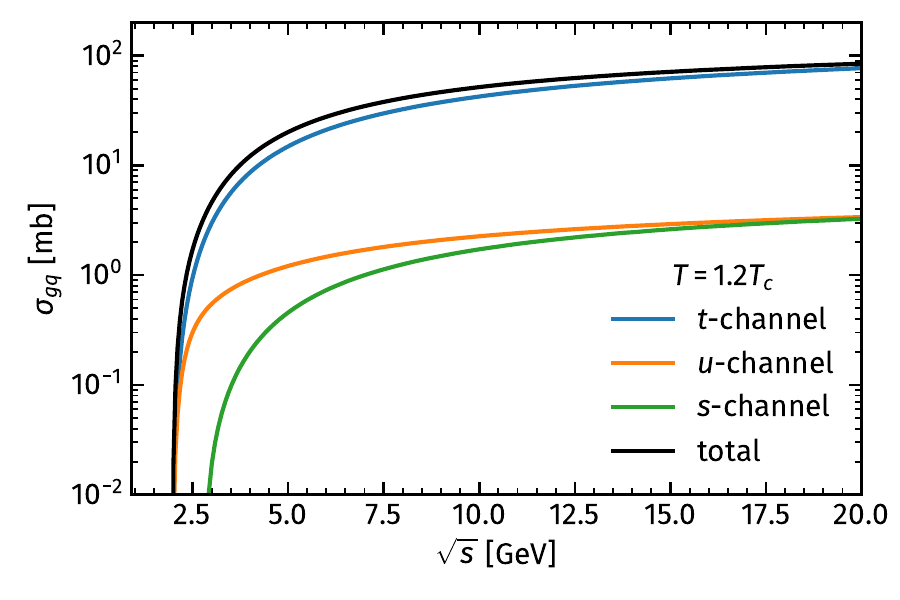}
  \caption{
    The contribution to the DQPM inelastic total cross sections for the $q+g \to q+g+g$ process at $T=1.2T_c$ from the $t$ channel (blue line), $u$ channel (orange line), $s$ channel (green line), and from all channels (black line).
  }
  \label{fig:XSi-gq_tusa}
\end{figure}

Figure \ref{fig:XSi-gq_tusa} illustrates the contribution to the DQPM inelastic total cross sections for the $q+g \to q+g+g$ process from different interaction channels: $t$ channel (blue line), $u$ channel (orange line), $s$ channel (green line), and from all channels (black line). As it is seen from the figure, the contributions from the $u$ and $s$ channels are almost two orders of magnitude lower than for the $t$ channel and do not give a significant contribution to the cross section for the entire energy region. Thus we neglect the contributions from the $u$ and $s$ channels in the further calculations (similar to the $q+q \to q+q+g$ process).


\subsection{Total inelastic cross section versus \texorpdfstring{$\srs$}{sqrt(s)} and \texorpdfstring{$T$}{T}}

\subsubsection{\texorpdfstring{$q+q \to q+q+g$}{q+q -> q+q+g} process}

\begin{figure}[ht!]
  \centering
  \includegraphics[width=\columnwidth]{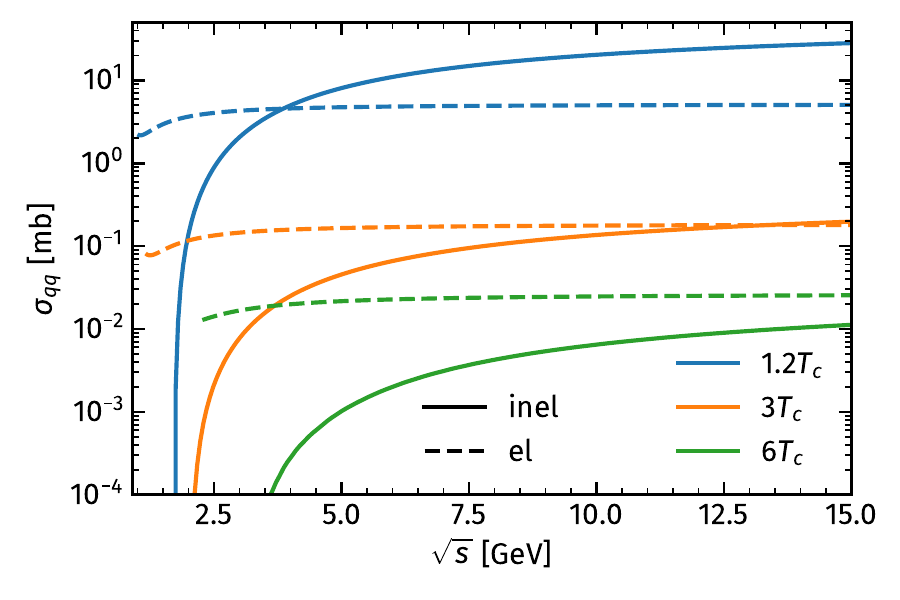}
  \caption{
    Total inelastic $q+q \to q+q+g$ cross sections (solid lines) as a function of the collision energy $\sqrt{s}$ at different temperatures $T =1.2T_c$ (blue), $T =3T_c$ (orange), and $T =6T_c$ (green) in comparison to the elastic $q+q \to q+q$ cross sections (dashed lines).
  }
  \label{fig:XSi-S}
\end{figure}

\begin{figure}[ht!]
  \centering
  \includegraphics[width=\columnwidth]{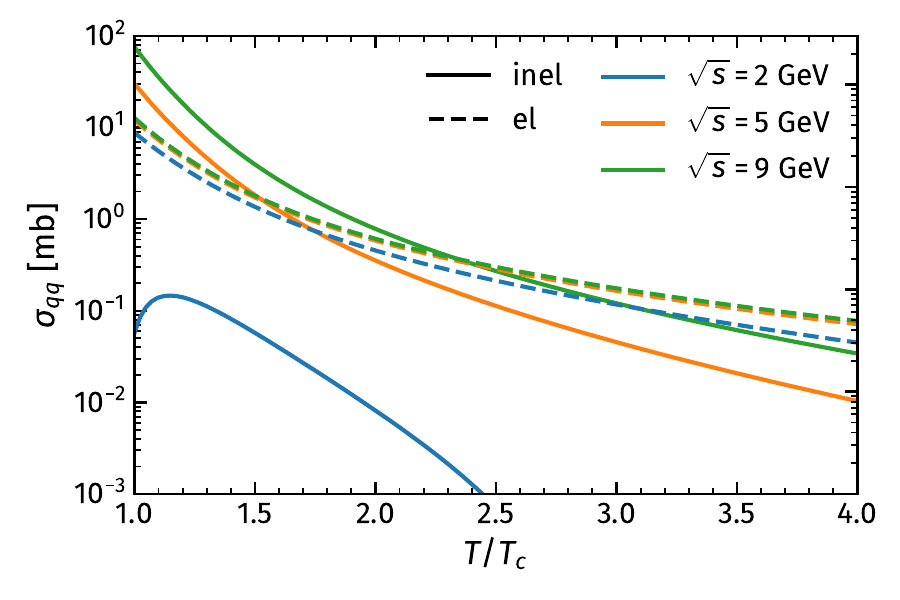}
  \caption{
    Total inelastic $q+q \to q+q+g$ cross sections (solid lines) as a function of temperature $T$ at different collision energies $\sqrt{s}=2$ GeV (blue), $\sqrt{s}=5$ GeV (orange) and $\sqrt{s}=9$ GeV (green) in comparison to the elastic $q+q \to q+q$ cross sections (dashed lines).
  }
  \label{fig:XSi-T}
\end{figure}
 
In Fig. \ref{fig:XSi-S} we present the DQPM results for the total cross sections for gluon radiation from $q+q \to q+q+g$ processes (solid lines) as a function of the collision energy $\sqrt{s}$ at different temperatures $T=1.2T_c$ (blue), $T=3T_c$ (orange), and $T=6T_c$ (green). The inelastic cross sections are compared to the elastic $q+q \to q+q$ cross sections (dashed lines) at the same $T$. Both processes are calculated for on-shell partons where their masses are taken as the pole masses of the spectral functions at a given temperature $T$. In Fig. \ref{fig:XSi-T} the total inelastic cross sections (solid lines) are shown as a function of temperature $T$ at the different collision energies $\sqrt{s}=2$ GeV (blue), $\sqrt{s}=5$ GeV (orange), and $\sqrt{s}=9$ GeV (green) in comparison to the elastic cross sections (dashed lines).

As one can see from the figures, both elastic and inelastic cross sections increase monotonically with energy and decrease with temperature. However, the form of these dependencies is rather different. For low energies the inelastic cross section is suppressed for the entire range of temperatures. At the same time for high temperatures the inelastic cross section is also suppressed for the range of energies up to 15 GeV. However, in the low-temperature regime the inelastic cross section can become significant. For example, for $T=1.2T_c$ the inelastic cross section is of the same order as the elastic one already at $\srs \approx 3.5$ GeV and $T=3T_c$ at $\srs \approx 12$ GeV. 

The reason for such a behavior of the inelastic cross section lies in the form of the scattering amplitudes, which for the inelastic reactions [Eqs. (\ref{eq:M})] are proportional to the coupling cubed $(\Msq{2\to 3} \propto \alpha_s^3)$, while for the elastic reaction the amplitudes are proportional to the coupling squared $(\Msq{2\to 2} \propto \alpha_s^2)$. Since in the DQPM the coupling constant is temperature dependent, $\alpha_s(T)=g^2(T)/(4\pi)$, and it grows above 1 in the vicinity of the critical temperature (cf. Fig. \ref{fig:alphas}), it leads to a rapid increase of the inelastic amplitude as the temperature decreases. Thus, the relative contribution of elastic and inelastic cross sections depends strongly on the temperature dependence of the coupling constant, $\alpha_s(T)$. However, if $\alpha_s$ is taken as a fixed constant below 1, e.g., $\alpha_s=0.3$ as in many pQCD calculations or effective models, the inelastic cross sections are suppressed.

\begin{figure}[t!]
  \centering
  \includegraphics[width=\columnwidth]{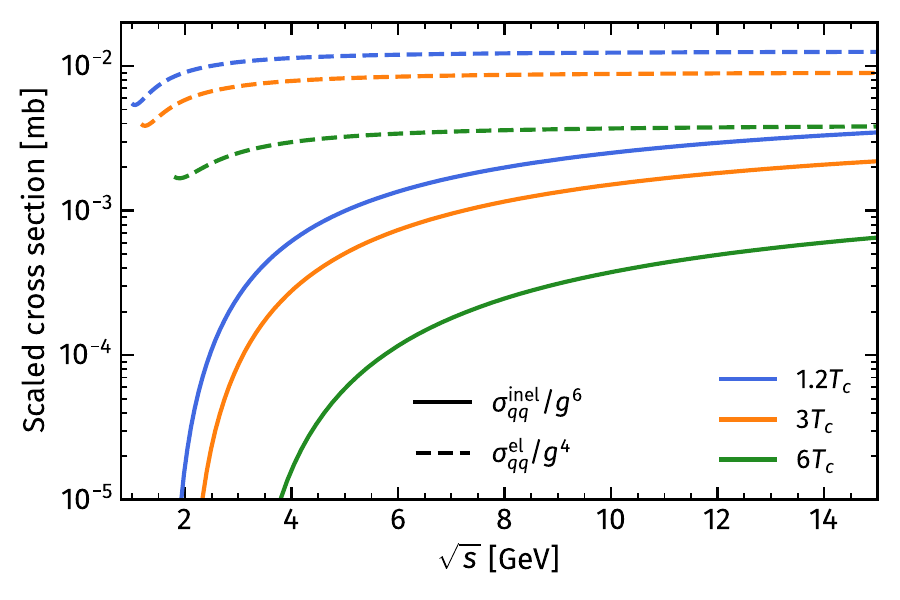}
  \caption{
    Total inelastic (solid lines) and elastic (dashed lines) cross sections scaled by $g^4$ and $g^6$, respectively, versus the invariant energy $\srs$.
  }
  \label{fig:XSi-S_scaled}
\end{figure}

This is demonstrated in Fig. \ref{fig:XSi-S_scaled}, which shows the total inelastic (solid lines) and elastic (dashed lines) cross sections scaled by $g^4(T)$ and $g^6(T)$, respectively. One can see a strong suppression of inelastic cross sections at low $\srs$ compared to elastic ones since the energy is not sufficient to emit a heavy thermal gluon. While $2\to 3$ cross sections grow with increasing energy, they are below the elastic $2\to 2$ cross sections by about of factor of 10 even at $\srs=15$ GeV for all $T$.

We note that the estimated averaged temperature of the QGP created in central Au+Au collisions at $\srs = 200$ GeV at RHIC is on a level of $1.5T_c$ and that in central Pb+Pb collisions at $\srs = 2.76$ TeV at LHC is about $2T_c$. For these temperatures the mean values of $\srs$ of colliding partons (for the DQPM) are less than 2 GeV, which according to the above figures implies that the processes of gluon radiation by thermal partons are expected to be negligible. However, these processes may play an important role in the case of jets (fast light partons or heavy quarks) as they can have a very large momentum (above 10 GeV/$c$) and therefore produce collisions with significant values of $\srs$.


\subsubsection{\texorpdfstring{$q+g \to q+g+g$}{q+g -> q+g+g} process}

\begin{figure}[ht!]
  \centering
  \includegraphics[width=\columnwidth]{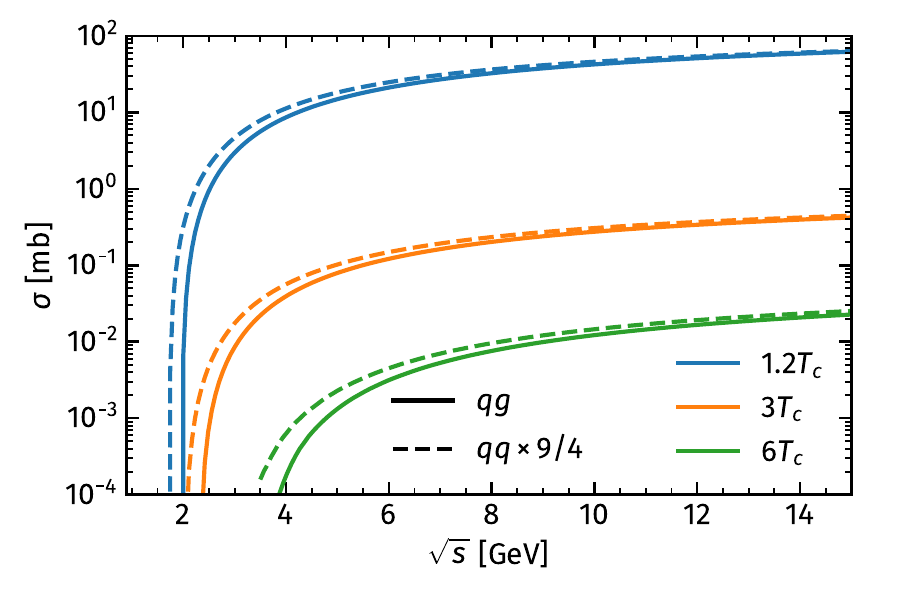}
  \includegraphics[width=\columnwidth]{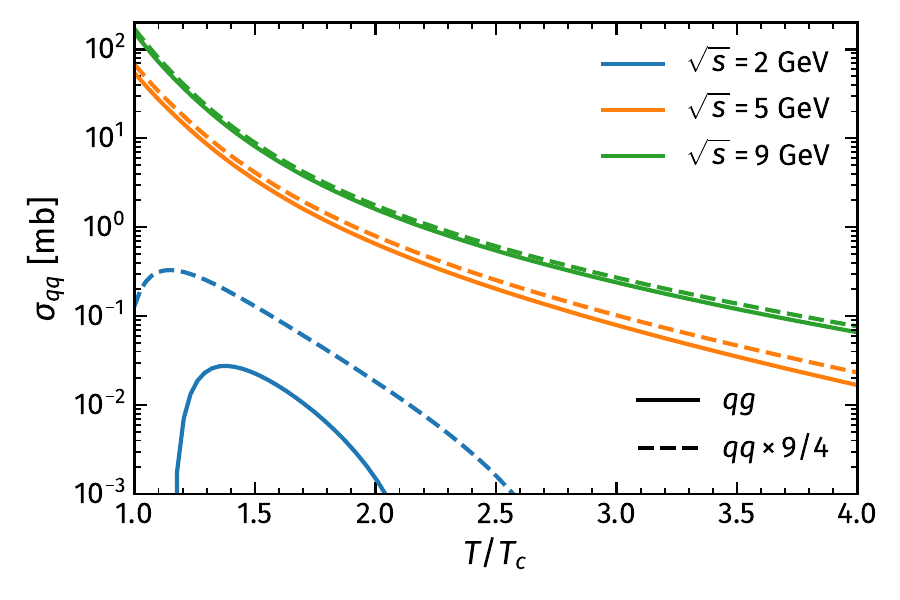}
  \caption{
    Comparison of the total inelastic cross sections of the $q+g \to q+g+g$ reaction with the inelastic $q+q \to q+q+g$ reaction (scaled by the factor of 9/4) as a function of the collision energy $\srs$ for different temperatures $T=1.2T_c,\ 3T_c$, and $6T_c$ (upper plot) and as a function of the temperature for different collision energies $\srs=2,\ 5$, and $9$ GeV (lower plot).
  }
  \label{fig:XSi_qq-vs-qg}
\end{figure}

Figure \ref{fig:XSi_qq-vs-qg} shows the total inelastic cross sections for $q+g \to q+g+g$ reactions -- calculated by summing the diagrams in Fig. \ref{fig:diags_gq} according to the amplitudes (\ref{eq:Mglue}) -- as a function of the collision energy $\srs$ for different medium temperatures $T=1.2T_c,\ 3T_c$, and $6T_c$ (upper plot) and as a function of $T$ for different $\srs=2,\ 5$, and $9$ GeV (lower plot). The $q+g \to q+g+g$ cross sections are compared to the $q+q \to q+q+g$ cross section scaled by a factor of 9/4. The ratio for the $q+q$ and $q+g$ cross sections reaches the expected pQCD color factor limit $4/9$ \cite{Berends:1981rb} at high energies. At low energies the inelastic $q+g$ cross sections are smaller than the $q+q$ ones due to the different thresholds for $q+q$ and $q+g$ reactions, i.e., $\srs_{th}^{q+q}=2m_q+m_g < \srs_{th}^{q+g}=m_q+2m_g$, since the calculations are carried out for the pole masses of initial partons and the pole gluon mass is larger than the pole quark mass.


\subsection{\texorpdfstring{$\mu_B$}{muB} dependence of the gluon radiative cross sections}

\begin{figure}[ht!]
  \centering
  \includegraphics[width=\columnwidth]{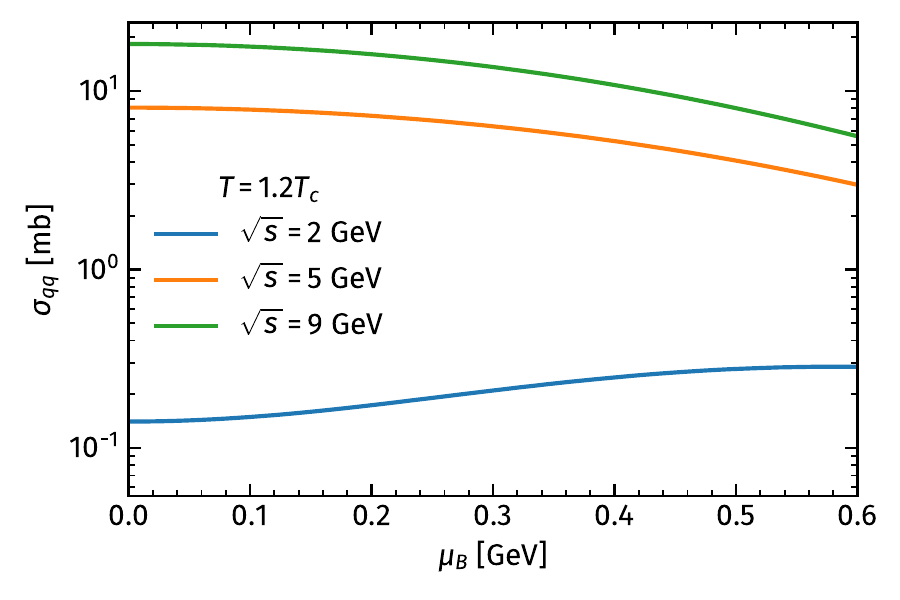}
  \includegraphics[width=\columnwidth]{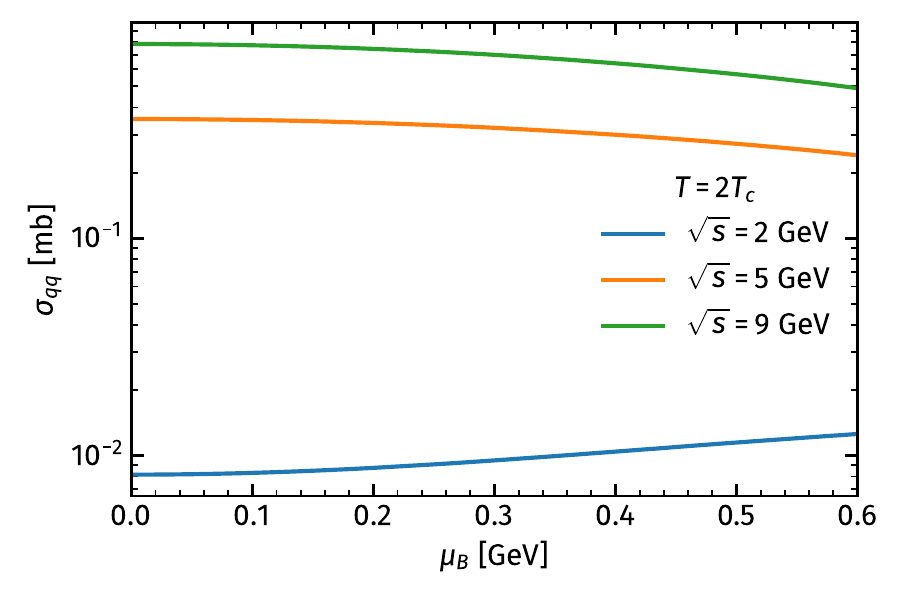}
  \caption{
    Total inelastic $q+q \to q+q+g$ cross section as a function of the baryon chemical potential $\mu_B$ for $T=1.2T_c$ (upper plot) and $T=2T_c$ (lower plot) at different collision energies $\sqrt{s}=2$ GeV (blue), $\sqrt{s}=5$ GeV (orange), and $\sqrt{s}=9$ GeV (green) for $T_c = 0.158$ GeV.
  }
  \label{fig:XSi_muB}
\end{figure}

Figure \ref{fig:XSi_muB} shows the total inelastic $q+q \to q+q+g$ cross section as a function of the baryon chemical potential $\mu_B$ for $T=2T_c$ at different collision energies $\sqrt{s}=2$ GeV (blue), $\sqrt{s}=5$ GeV (orange), and $\sqrt{s}=9$ GeV (green). One can see that the $\mu_B$ dependence of the cross section $\sigma_{qq}$ is not strong; however, it is nontrivial and changes with $T$ and $\srs$. We note that in the DQPM the quark and gluon masses as well as $\alpha_s$ decrease with increasing $\mu_B$ \cite{Moreau:2019vhw,Soloveva:2020hpr}. At low temperature $T=1.2T_c$ the cross section $\sigma_{qq}$ increases for $\sqrt{s}=2$ GeV due to the lowering of the threshold $\srs_{th}^{q+q}=m_q+m_q+m_g$ for gluon production in the $2\to 3$ reaction which increases the open phase space. With increasing energy ($\sqrt{s}=5, \ 9 $ GeV) the threshold effect does not play a role anymore since the open phase space is large anyhow and the cross section decreases with increasing $\mu_B$ partially due to the reduction of $\alpha_s(\mu_B)$. Similarly, at $T=2T_c$ the cross section (which substantially decreases with $T$; cf. Fig. \ref{fig:XSi-T}) decreases with increasing $\mu_B$. 


\subsection{Interaction rates and relaxation times}

The thermal interaction rate and relaxation time of the on-shell and off-shell partons within the DQPM have been studied in Refs. \cite{Moreau:2019vhw,Soloveva:2020hpr} by accounting for the elastic $q+q$ and $q+g$ reactions. Here we investigate the contribution of gluon radiative reactions to the interaction rate and relaxation time. We note that in this study we consider the on-shell case for elastic and inelastic scattering -- since we concentrate on the relative contributions of these reactions -- to reduce the numerical complexity of the full off-shell calculations for $2\to 3$ processes. However, we point out that the off-shell elastic rate evaluated earlier in Refs \cite{Moreau:2019vhw,Soloveva:2020hpr} is smaller than the on-shell rate here due to the averaging over the spectral functions of final partons. 

We recall that the interaction rate for the corresponding on-shell partons (i.e., taken at their pole masses) is given by
\begin{align}
  \label{eq:IR}
  \Gamma_j^{\text{inel}}(p_j,T,\mu_q) = \frac{1}{2E_j}\sum_{i=q,\bar{q},g}
   \int\frac{d^3p_i}{\dpi^3 2E_i} d_i f_i \int\frac{d^3p_1}{\dpi^3 2E_1}
  \nonumber\\
  \times \int\frac{d^3p_2}{\dpi^3 2E_2} \int\frac{d^3p_3}{\dpi^3 2E_3} 
  (1 \pm f_1)(1 \pm f_2)(1 \pm f_3)
  \nonumber\\
  \times \ \Msq\ (2\pi)^4 \delta^{(4)}(p_j + p_i - p_1 - p_2 - p_3),
\end{align}
where $d_i$ is the degeneracy factor for spin and color [$2N_c$ for quarks and $2(N_c^2-1)$ for gluons], $f_i = f_i(E_i,T,\mu_q)$ are the Fermi distribution functions for quarks, and $f_i = f_i(E_i,T)$ are the Bose distribution functions for gluons. Here $v_{\text{rel}}=F/(4 E_{\text{j}} E_i)$ with $F$ being the flux of the incident particles. The Pauli-blocking $(-)$ and Bose-enhancement $(+)$ factors account for the available density of final states.

To evaluate the average width of the parton $j$, we finally have to average its interaction rate \eqref{eq:IR} over its momentum distribution,
\begin{equation}
  \Gamma_j^{\text{inel}}(T,\mu_q) = \frac{d_j}{n_j(T,\mu_q)} \int \frac{d^3 p_j}{\dpi^3} f_j(E_j,T,\mu_q) \Gamma_j(p_j,T,\mu_q)
  \label{eq:GTmu}
\end{equation}
with the on-shell density of partons $j$ at $T$ and $\mu_q$ given by
\begin{equation}
\label{eq:f}
  n_j(T,\mu_q) = d_j \int \frac{d^3 p}{\dpi^3} f_j(E_j,T,\mu_q).
\end{equation}

\begin{figure}[t!]
  \centering
  \includegraphics[width=\columnwidth]{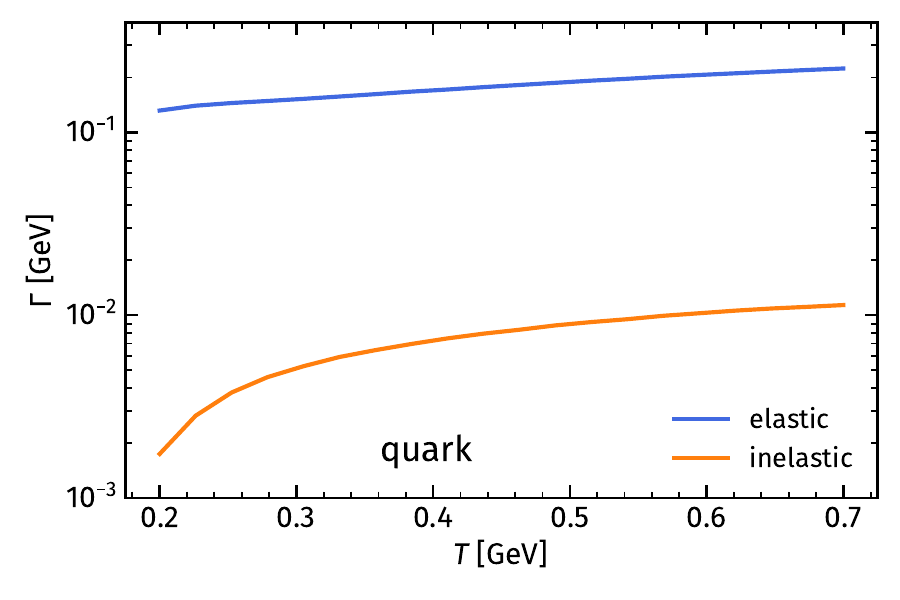}
  \includegraphics[width=\columnwidth]{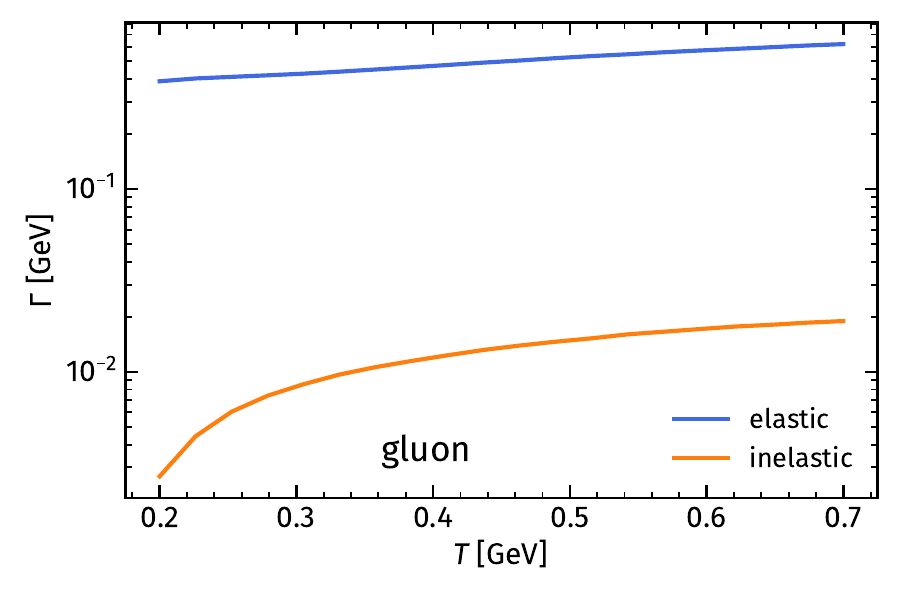}
  \caption{On-shell interaction rate $\Gamma$ of a light quark (upper plot) and a gluon (lower plot) as a function of the temperature $T$ for $\mu_q = 0$ for elastic (blue lines) and for inelastic (orange lines) processes.}
  \label{fig:inter_rate}
\end{figure}

Figure \ref{fig:inter_rate} illustrates the on-shell interaction rate of a light quark (upper plot) and for the gluon (lower plot) as a function of the temperature for elastic (blue line) and inelastic (orange line) processes. Here, similar to Refs. \cite{Moreau:2019vhw,Soloveva:2020hpr}, for the elastic light quark scattering (e.g., $u+u$) we account for the contributions of all the possible interaction channels such as
 $uu \to uu$,
 $u\ubar \to u\ubar$,
 $u\ubar \to d\dbar$,
 $u\ubar \to s\sbar$,
 $ud \to ud$,
 $u\dbar \to u\dbar$,
 $us \to us$,
 $u\sbar \to u\sbar$, and
 $ug \to ug$.
For the inelastic scatterings we account for the same channels as for elastic (with an additional emitted gluon in the final state); however, we omit the contributions from the annihilation processes such as $u\ubar \to d\dbar$ and $u\ubar \to s\sbar$ and assume identical contributions from all $uq \to uq$ processes for the reason described in Sec. \ref{sec:ampl_qq}.

For the gluon elastic scatterings the following reactions are considered (similar to Refs. \cite{Moreau:2019vhw,Soloveva:2020hpr}):
 $gu \to gu$,
 $g\ubar \to g\ubar$,
 $gd \to gd$,
 $d\dbar \to g\dbar$,
 $gs \to gs$,
 $g\sbar \to g\sbar$, and
 $gg \to gg$.
For the inelastic scatterings, we account for the same channels (with an additional emitted gluon in the final state). We note that the $gg\to ggg$ channel has been evaluated using the relation $\sigma_{gg} = \frac{9}{4}\sigma_{gq}$. We note that the evaluated on-shell interaction rate here is the same as in previous DQPM studies in Refs. \cite{Moreau:2019vhw,Soloveva:2020hpr}.

\begin{figure}[th!]
  \centering
  \includegraphics[width=\columnwidth]{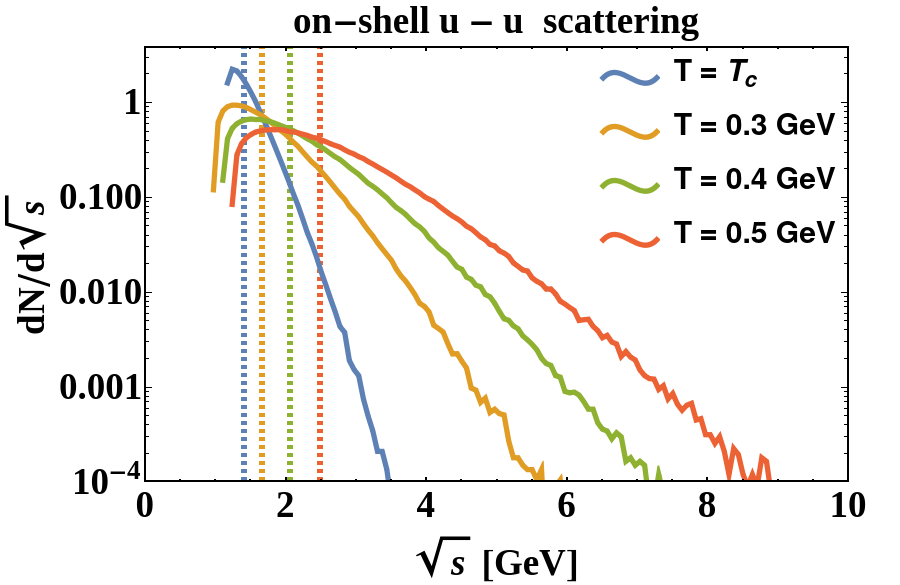}
  \caption{
    Distribution of on-shell $u$-quark scatterings as a function of the invariant energy $\sqrt{s}$ for different temperatures at $\mu_q = 0$. The vertical dashed lines correspond to the mean $\sqrt{s}$ for each temperature.
  }
  \label{fig:dens}
\end{figure}

As follows from Fig. \ref{fig:inter_rate}, the inelastic rates of thermal light quarks and gluons are strongly suppressed at all temperatures compared to the elastic rates in the QGP medium of quarks and gluons with thermal masses.
 
For a better understanding of this result we present in Fig. \ref{fig:dens} the distribution of on-shell $u$-quark scatterings (assuming the scattering probability to be equal to 1) as a function of the invariant energy $\sqrt{s}$ for different temperatures at $\mu_B = 0$. The vertical dashed lines correspond to the mean $\sqrt{s}$ for each temperature $T$. The sharp threshold is due to the large pole masses of scattered quarks (cf. Fig. \ref{fig:Masses}). We note that in the off-shell case -- as considered in Refs. \cite{Moreau:2019vhw,Soloveva:2020hpr} -- the distribution $dN/d\sqrt{s}$ goes to zero by accounting for all possible masses of the initial quarks according to their spectral functions.

As seen from Fig. \ref{fig:dens}, the $dN/d\srs$ distribution of scattered quasiparticles favors low values of $\srs$ due to the Fermi-Dirac distribution function $f_j(E_j,T,\mu_q)$ which suppresses collisions at large $\srs$ in the thermal QCD medium. Thus, only the low-$\srs$ region of elastic and inelastic cross sections dominantly contributes to the interaction rates $\Gamma$. As follows from Fig. \ref{fig:XSi-T}, at all achievable $\srs$ (cf. Fig. \ref{fig:dens}) the inelastic cross sections are much smaller than the elastic ones at all $T$ which leads to smaller inelastic rates compared to the elastic rates at all $T$. Thus, the contribution of radiative processes -- the emission of massive thermal gluons by collisions of massive partons -- is negligible for the thermal properties of the sQGP within the quasiparticle picture. The situation will change when considering the collisions of fast jet quarks with the thermal medium since the available $\srs$ is large in this case (cf. Fig. \ref{fig:XSi-S}); thus the radiative contribution becomes important \cite{Grishmanovskii:2023new}.

\begin{figure}[t!]
  \centering
  \includegraphics[width=\columnwidth]{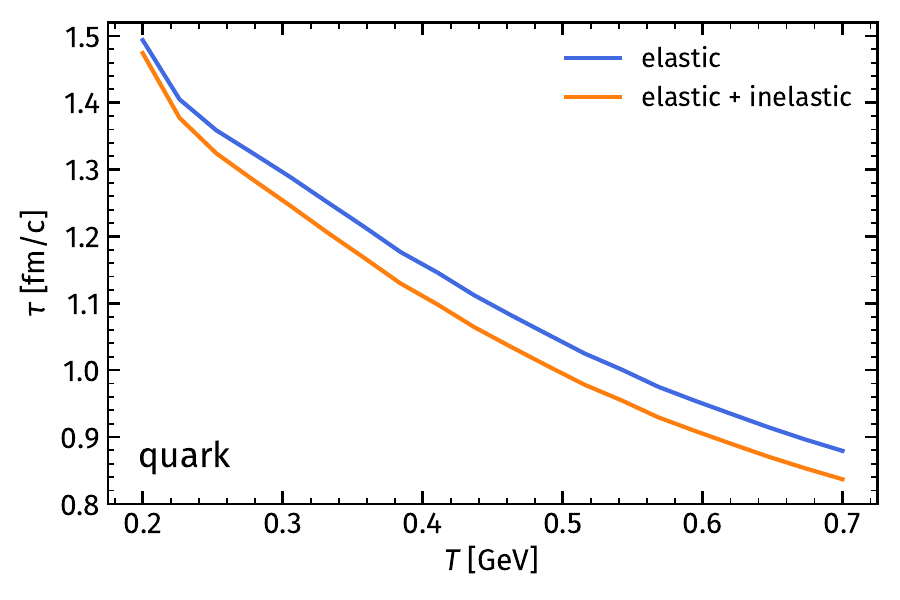}
  \includegraphics[width=\columnwidth]{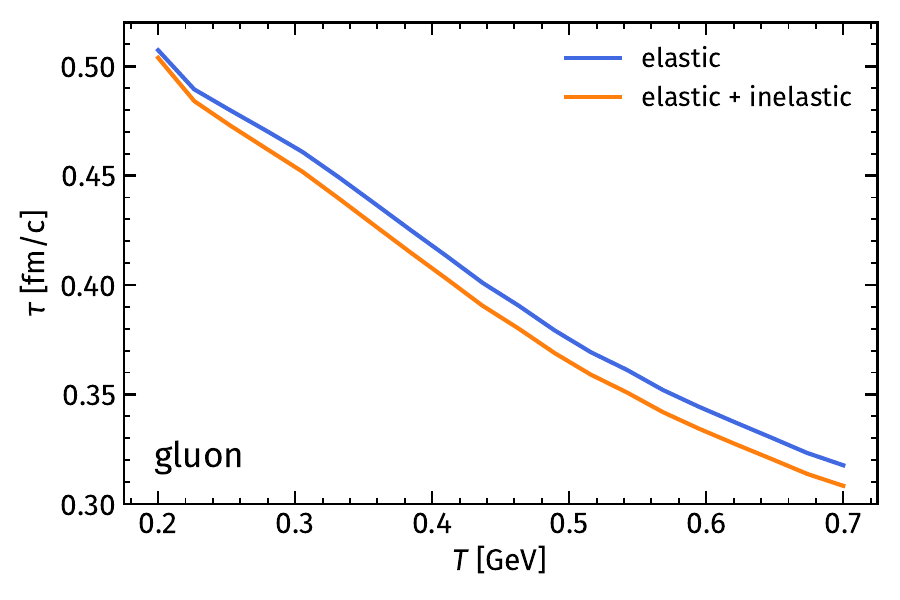}
  \caption{
    Relaxation time $\tau$ of a light quark (upper plot) and gluon (lower plot) as a function of the temperature $T$ for $\mu_q = 0$ evaluated by the average parton interaction rate for elastic contribution only (blue lines) and for the sum of elastic and inelastic contributions (orange lines).
  }
  \label{fig:tau}
\end{figure}

The interaction rate can be directly related to the relaxation time $\tau = \hbar c/\Gamma$ which is shown in Fig. \ref{fig:tau}. There $\tau$ for a light quark (upper plot) and gluon (lower plot) is presented as a function of the temperature $T$ evaluated by the average parton interaction rate for elastic contributions only (blue line) and for the sum of elastic and inelastic contributions (orange line) from Fig. \ref{fig:inter_rate}. One can see that by accounting for inelastic processes the relaxation time of thermal sQGP partons is only slightly shortened. As discussed above this is related to the fact that the thermal radiation of heavy gluons from $q+q$ and $q+g$ scattering is suppressed compared to their elastic scattering.

Since the inelastic reactions only slightly contribute to the relaxation times of quarks and gluons, the DQPM transport coefficients of the thermal sQGP -- such as the specific shear and bulk viscosities $\eta/s$ and $\zeta/s$, electric conductivity $\sigma_Q/T$, as well as the diffusion coefficient matrix -- evaluated for the elastic reactions within the relaxation time approximation (RTA) \cite{Moreau:2019vhw,Soloveva:2019xph,Fotakis:2021diq} will change insignificantly. Only in the high-temperature region are transport coefficients from the RTA including the inelastic contributions expected to be smaller when compared to previous estimates that solely considered elastic ones.


\section{Summary}
\label{sec:summary}

We have studied the inelastic versus elastic scatterings in the sQGP based on the effective DQPM, which describes the nonperturbative nature of the sQGP at finite temperature $T$ and baryon chemical potential $\mu_B$ in terms of off-shell quarks and gluons with properties (masses and widths) adjusted in order to reproduce the lQCD equation of state (EoS) for the QGP in thermodynamic equilibrium. We stress that, contrary to the pQCD-based models with massless weakly interacting quarks and gluons, the concept of massive strongly interacting quasiparticles in the DQPM allows not only to reproduce the lQCD EOS at zero and finite $\mu_B$, but to explore QCD in the nonperturbative regime of the ($T,\mu_B)$ phase diagram where lQCD results are not yet available; however, this region is relevant for heavy-ion physics. While in previous DQPM studies we focused on elastic scatterings of quasiparticles (which are assumed to be dominant in the quasiparticle picture due to the large thermal masses) by evaluating the Born-type scattering diagrams with effective propagators and vertices, here we extend our study to the evaluation of inelastic processes with the goal to quantify explicitly the relative contribution of the emission of massive gluons from scatterings of thermal quarks and gluons to the thermal properties of the QGP.

We summarize our study as follows:
\begin{itemize}
\item We have calculated the thermal gluon radiation processes from the massive quark-quark ($q+q \to q+q+g$) and quark-gluon ($q+g \to q+g+g$) scatterings (similar for the scatterings including antiquarks) by evaluating explicitly the leading-order Feynman diagrams for $2\to 3$ processes with effective propagators and vertices from the DQPM for the first time without any further approximations and simplifications. The calculations of $2\to 3$ cross sections have been performed for the \textquote{on-shell} case, where the masses of scattered initial as well as final partons, including the emitted gluon, have been taken at their pole masses, thus omitting the integrations over the spectral functions of final partons as  for the off-shell calculation of elastic scattering in earlier DQPM studies \cite{Moreau:2019vhw,Soloveva:2020hpr}. 
\item We have shown that for $q+q$ and $q+g$ radiative cross sections the $t$ channel is the dominant channel compared to $u$ and $s$ channels. Moreover, a large contribution to the $q+q\to q+q+g$ cross sections stems from diagram 5 in Fig. \ref{fig:diags_qq} with the emission of a gluon from the gluon exchange lines; this contribution is larger than from each of the other diagrams 1-4 (they give about equal contributions) at all $\srs$. However, the total cross section is smaller than the sum of the squared amplitudes and even the squared amplitude of diagram 5 due to a strong destructive interference between different amplitudes, which shows the importance of an explicit evaluation of all diagrams without approximations.
\item We have found that the DQPM radiative cross sections for $q+q$ and $q+g$ reactions scale with a factor $4/9$ at high collision energies $\srs$, in line with the pQCD calculations. However, at low $\srs$ the cross sections for $q+g\to q+g+g$ reactions are smaller than for $q+q\to q+q+g$ reactions due to the larger threshold (i.e., the pole mass of thermal gluons is larger than the pole mass of quarks).
\item We have proven that the DQPM results for $2\to 3$ processes reproduce the pQCD calculations in the limit of zero masses and widths of quasiparticles in the DQPM. In particular, we have shown the identity of the DQPM and pQCD amplitudes \cite{Berends:1981rb,Ellis:1985er} in the \textquote{pQCD} limit.
\item Moreover, we have compared our results for the angular distributions of the emitted gluons and final scattered quarks from $q+q \to q+q+g$ processes to the pQCD-based BAMPS model calculations and have found a qualitative agreement. The quantitative differences between the DQPM and the pQCD BAMPS results can be attributed to the following reasons: (i) the thermal masses of scattered quarks and emitted gluons in the DQPM versus zero mass partons in BAMPS; (ii) differences in the propagators -- the necessity to introduce a Debye mass (i.e., an extra parameter in the model) to regularize the infrared divergence in the gluon pQCD propagator while the \textquote{resummed} DQPM propagator has no divergence due to the self-consistent dynamical widths of the 2PI approach; and (iii) different strong-coupling constant $\alpha_s$ used in the calculations -- while the pQCD calculations assume a small $\alpha_s$ (e.g. in our study $\alpha_s=0.3$), independent of temperature, the DQPM calculations involve $T$-dependent $\alpha_s(T)$ which grows above 1 for $T\to T_c$ in line with the lQCD data (cf. Fig. \ref{fig:alphas}).
\item We have calculated the energy and temperature dependencies of the total and differential radiative cross sections for the thermal quark-quark and quark-gluon scatterings and compared them to the corresponding elastic cross sections. We have found that the elastic cross sections dominate the radiative ones at low-$\srs$ energies for $T \geq 2T_C$; however, at low $T\to T_c$ the radiative cross sections become larger. The reason is related to the fact that the inelastic cross sections are proportional to the coupling cubed $(\sigma_{2\to 3} \propto \alpha_s^3(T))$ while elastic ones are proportional to the coupling squared $(\sigma_{2\to 2} \propto \alpha_s^2(T))$. 
\item We have studied the $\mu_B$ dependence of the radiative cross sections and found a relatively weak dependence on $\mu_B$: at larger $\srs$, where the phase space is open sufficiently, the total cross sections decrease with increasing $\mu_B$.
\item We have estimated the interaction rate $\Gamma$ and relaxation time $\tau$ of quarks and gluons in the sQGP versus the temperature $T$. We have found that the transition rate defined by elastic scattering strongly dominates the inelastic transition rate. Thus, the relaxation time $\tau$ calculated within the elastic plus inelastic transition rates $\Gamma$ is only slightly shorter than that with elastic $\Gamma$ only. It implies that the radiative processes -- with the emission of massive gluons -- are strongly suppressed in the nonperturbative sQGP medium within the dynamical quasiparticle model.
\end{itemize}

We mention that our study of thermal radiative processes can be extended for the investigation of jet (highly energetic partons) attenuation in the sQGP and jet transport properties. This work is in progress \cite{Grishmanovskii:2023new}.


\section*{ACKNOWLEDGEMENTS}

The authors acknowledge inspiring discussions with J. Aichelin, W. Cassing, M. Djordjevic, P.-B. Gossiaux and O. Kaczmarek. Furthermore, we acknowledge support by the Deutsche Forschungsgemeinschaft (DFG, German Research Foundation) through the grant CRC-TR 211 \textquote{Strong-interaction matter under extreme conditions} -- project number 315477589 -- TRR 211. I.G. also acknowledges support from the \textquote{Helmholtz Graduate School for Heavy Ion Research}. This work is supported by the European Union's Horizon 2020 research and innovation program under Grant Agreement No. 824093 (STRONG-2020). The computational resources have been provided by the Goethe-HLR Center for Scientific Computing.


\appendix
\begin{widetext}
\section{Gluon polarization}
\label{appx:gluon_polarization}
 
In the case of a massive gluon, the sum over polarizations is given by
\begin{equation}
  \sum_{\text{pol.}} (\epsilon_i)_\mu (\epsilon^\star_i)_{\mu'} = - g_{\mu\mu'} + \frac{(k_i)_{\mu}(k_i)_{\mu'}}{(M_i)^2_g} ,
\label{eq:GP}
\end{equation}
where in addition to the first \textquote{simple} term -- as for massless gluons -- there is the second term which is over proportional to the gluon thermal mass.

When calculating the cross sections for gluon radiation, we have found that the first term gives the dominant contribution such that the total cross section calculated including both terms differs only very little from those calculated with the first term only. This is illustrated in Fig. \ref{fig:XSi-S_glupol} which shows the total $q+q \to q+q+g$ cross section as a function of the collision energy for the \textquote{full} (blue line) and \textquote{simple} (orange line) gluon polarization. The comparison shows that the second term in the gluon polarization sum (\ref{eq:GP}) gives an almost negligible contribution to the cross section and therefore can be omitted in all further calculations.

\begin{figure}[ht!]
  \centering
  \includegraphics[width=0.55\columnwidth]{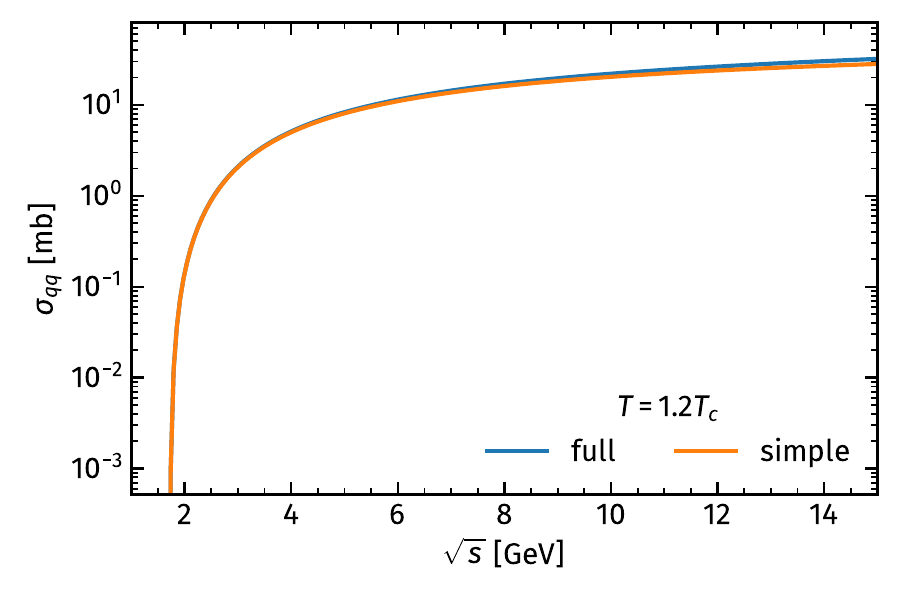}
  \caption{
    Total $q+q \to q+q+g$ cross section at $T=1.2T_c$ as a function of the collision energy for the \textquote{full} (blue line) and \textquote{simple} (orange line) gluon polarizations.
  }
  \label{fig:XSi-S_glupol}
\end{figure}


\section{Details on the calculation of the cross section}
\label{appx:cross_section}

The differential cross section for a $2\to 3$ process for on-shell particles $(a+b \to 1+2+3)$ is given by
\begin{equation}
  d\sigma_{2\to 3} = \frac{\Msq{}}{F}d\Phi_3,
\end{equation}
where $d\Phi_3$ denotes the three-body Lorentz invariant phase-space for on-shell particles:
\begin{equation}
  d\Phi_3 = \dpi^4 \frac{d^3 p_1}{\dpi^3 2E_1} \frac{d^3 p_2}{\dpi^3 2E_2} \frac{d^3 p_3}{\dpi^3 2E_3} \delta^4(p_a + p_b - p_1 - p_2 - p_3).
  \label{eq:lips_general}
\end{equation}

Note, that in general, the total cross section is not only a function of the system energy but also a function of all parton masses and widths, i.e.
\begin{equation}
  \sigma_{2\to3} = \sigma_{2\to3}(s,m_a,m_b,m_1,m_2,m_3,M_q,M_g,\gamma_q,\gamma_g)
\end{equation}

Obtaining this total cross section implies, in general, a nine-dimensional integration, where four integrations can be done analytically using the properties of the $\delta$ function. There are several ways to perform this integration depending on the choice of integration variables.


\subsection{Variant 1}
\label{appx:var1}

\begin{figure}[ht!]
  \label{fig:2to3_schematic}
  \includegraphics[width=0.6\columnwidth]{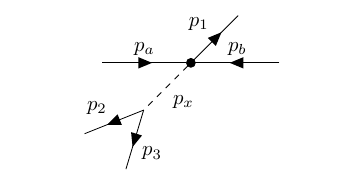}
  \caption{Schematic view of a $2\to3$ reaction. }
\end{figure}

The first way to perform the integration over the $\delta$ function is to consider a $2 \to 3$ reaction as $(2 \to 2) + (1 \to 2)$ by introducing the intermediate state $p_x$ (Fig. \ref{fig:2to3_schematic}), where
\begin{equation}
  p_x = p_2 + p_3, \quad p_x^2 = m_x^2, \quad m_2+m_3 < m_x < \sqrt{s} - m_1.
\end{equation}
Employing for $p_3$ the general relation
\begin{equation}
  \int \frac{d^3 p}{2E} = \int d^4 p \; \delta (p^2 - m^2) \, \theta(E) ,
  \label{eq:d3p_to_d4p}
\end{equation}
this stems from the property of the $\delta$ function,
\begin{equation}
  \delta(f(x)) = \sum_i \frac{1}{|f'(x_i)|}\delta(x-x_i),
  \label{eq:deltafunc_f}
\end{equation}
where $x_i$ are the roots of $f(x)$. Then after substituting 
\begin{equation}
  1 = \int d^4 p_x \delta^4(p_x - p_2 - p_3)
\end{equation}
into Eq. \eqref{eq:lips_general} and integrating over $p_3$, we obtain
\begin{equation}
  d\Phi_3 = \dpi^{-5} \frac{d^3 p_1}{2E_1} \frac{d^3 p_2}{2E_2} \; d^4 p_x \; \delta\left((p_x-p_2)^2-m_3^2\right) \; \delta^4(p_a+p_b-p_1-p_x).
\end{equation}
Introducing the same procedure for $p_1$ we arrive at
\begin{equation}
  d\Phi_3 = \dpi^{-5} \frac{d^3 p_2}{2E_2} d^4 p_x \; \delta\left((p_a+p_b-p_x)^2 - m_1^2\right) \; \delta\left((p_x-p_2)^2-m_3^2\right).
  \label{eq:phase-space_2}
\end{equation}
In the c.m. frame we now can simplify an expression inside the first $\delta$ function:
\begin{equation}
  (p_a+p_b-p_x)^2 = (p_a+p_b)^2 - 2(p_a+p_b)p_x + p_x^2 = s - 2\sqrt{s}E^C_x + m_x^2,
\end{equation}
where the upper index $C$ denotes the c.m. frame. Having in mind that
\begin{equation}
  d^4p = d^3pdE = p^2dpd\Omega dE = dm^2 \frac{p}{2}dE d\Omega,
\end{equation}
we now can write
\begin{equation}
  d^4p_x \delta\left((p_a+p_b-p_x)^2 - m_1^2\right) = dp_x^2 \frac{p_x^C}{2}dE_x^C d\Omega_x^C \; \delta(s - 2\sqrt{s}E^C_x + m_x^2 - m_1^2) = \frac{p_x^C}{2} \frac{dm_x d\Omega_x^C}{2\sqrt{s}} dE_x^C \delta(E_x^C - E_{x0}^C).
\end{equation}
For the second $\delta$ function in the rest frame of (2+3), we have
\begin{equation}
  (p_x-p_2) = p_x^2 - 2p_x p_2 + p_2^2 = m_x^2 - 2m_x^X E_2^X + m_2^2,
\end{equation}
where the upper index $X$ denotes the rest frame of (2+3). Substituting this into Eq. \eqref{eq:phase-space_2} we have
\begin{equation}
  \frac{d^3 p_2}{2E_2} \delta\left((p_x-p_x)^2 - m_3^2\right) = \frac{p_2^X}{2} dE_2^X d\Omega_2^X \delta(p_x^2 - 2E_x^X E_2^X + m_2^2 - m_3^2) = \frac{p_2^x}{2} \frac{d\Omega_2^X}{2E_x^X} dE_2^X \delta(E_2^X - E_{20}^X).
\end{equation}
Integrating over the $\delta$ functions we end up with
\begin{equation}
  d\Phi_3 = \dpi^{-5} \left(\frac{p_2^X}{4E_x^X} d\Omega_2^X\right) \left(\frac{p_x^C}{4\sqrt{s}} d\Omega_x^C\right) dm_x^2,
\end{equation}
and the expression for the differential cross section now reads
\begin{equation}
  d\sigma_{2\to3} = \frac{\Msq{2\to3}}{F} \dpi^{-5} \left(\frac{p_2^X}{4E_x^X} d\Omega_2^X\right) \left(\frac{p_x^C}{4\sqrt{s}} d\Omega_x^C\right) dm_x^2.
\end{equation}
Noting
\begin{equation}
  d\Omega_x^C = d\cos\theta d\varphi , \quad d\Omega_2^X = d\cos\xi d\psi, \quad E_x^X = m_x,
\end{equation}
we can now express the final integral for the total cross section:
\begin{equation}
  \sigma_{2\to3} = \frac{1}{8 \dpi^5 \sqrt{s} F} \int_0^1 d\cos\theta \int_0^{2\pi}d\varphi \int_0^1 d\cos\xi \int_0^{2\pi}d\psi \int_{m_2+m_3}^{\sqrt{s}-m_1} dm_x p_x^C p_2^X \Msq{2\to3}.
\end{equation}
The integration over the polar angle (of $p_x^C$) can be done analytically, giving a factor of $2\pi$. In order to obtain $\Msq{}$ one needs to reconstruct a full set of 4-momenta according to $\theta,\varphi,\xi,\psi,m_x$. Since $p_2^X$ and $p_x^C$ are expressed in different systems, one should perform the Lorentz transformation on one chosen system. For example, one can define the following set of momenta:
\begin{align}
  \sld{p}_a^C &= (E_a,0,0,p_a^C), \quad p_a^C = \lambda(\sqrt{s},m_a,m_b)/(2\sqrt{s}), \nonumber\\
  \sld{p}_b^C &= (E_b,0,0,-p_a^C), \nonumber\\
  \sld{p}_1^C &= (E_1,p_x^C \sin\theta \cos\varphi, p_x^C \sin\theta \sin\varphi, p_x^C \cos\theta), \quad p_x^C = \lambda(\sqrt{s},m_1,m_x)/(2\sqrt{s}), \nonumber\\
  \sld{p}_x^C &= (E_x, -\vec{p}_1), \nonumber\\
  \sld{p}_2^X &= (E_2^X, p_2^X \sin\xi \cos\psi, p_2^X \sin\xi \sin\psi, p_2^X \cos\xi), \quad p_2^X = \lambda(m_x,m_2,m_3)/(2m_x), \nonumber\\
  \sld{p}_3^X &= (E_3^X, -\vec{p}_2^X), \nonumber\\
  \sld{p}_2^C &= L_{-\vec{v}_x}(p_2^X), \nonumber\\
  \sld{p}_3^C &= L_{-\vec{v}_x}(p_3^X),
\end{align}
where $L_{\vec{v}}(p)$ denotes a Lorentz transformation of the 4-vector $p$ with the velocity $\vec{v}$ and
\begin{equation}
  \lambda(a,b,c) = \sqrt{\left(a^2-(b+c)^2\right)\left(a^2-(b-c)^2\right)}, \quad \vec{v}=\vec{p}_x/E_x.
\end{equation}


\subsection{Variant 2}
\label{appx:var2}

To properly compare DQPM results with the well-known Gunion-Bertsch approximation, we should express the integration in terms of another set of variables such as $(q_t,k_t,y,\phi)$, where $q_t$ is the transverse momentum transfer, $k_t$ is the transverse momentum of the emitted gluon, $y$ is the rapidity of the emitted gluon and $\phi$ is the angle between $k_t$ and $q_t$. For a given particle with 4-momentum $(E,p_x,p_y,p_z)$ we can express the components in terms of transverse momentum, rapidity, and azimuthal angle using the following relation:
\begin{equation}
  (E,p_x,p_y,p_z) = (m_t \cosh{y}, p_t \cos{\varphi}, p_t \sin{\varphi}, m_t \sinh{y}), \quad m_t = \sqrt{m^2 + p_t^2}.
\end{equation}
The scalar product of two 4-vectors is expressed now as
\begin{equation}
  p_i p_j = m_{i,t} m_{j,t}(\cosh{y_i} \cosh{y_2} - \sinh{y_1} \sinh{y_2}) - p_{i,t}p_{j,t}\cos \varphi.
\end{equation}

Starting again from the definition of the phase space from Eq. \eqref{eq:lips_general} and integration over $p_2$ using relations \eqref{eq:d3p_to_d4p} and \eqref{eq:deltafunc_f} we obtain
\begin{equation}
  d\Phi_3 = \frac{1}{128 \pi^5} \frac{d^3 p_1}{E_1}\frac{d^3 p_3}{E_3} \delta((p_a+p_b-p_1-p_3)^2 - m_2^2).
\end{equation}
Denoting the argument of the $\delta$ function $F=(p_a+p_b-p_1-p_3)^2 - m_2^2$ we have
\begin{align}
  F &= \left((p_a + p_b) - (p_1+p_3)\right)^2 - m_2^2 = (p_a + p_b)^2 - 2(p_a + p_b)(p_1 + p_3) + (p_1+p_3)^2 - m_2^2 \nonumber \\
  &= s - 2(E_a+E_b)(E_1+E_3) + 2(\vec{p}_a+\vec{p}_b)(\vec{p}_1+\vec{p}_3) + m_1^2 + m_3^2 + 2p_1 p_3 - m_2^2. 
\end{align}
In the c.m. frame $\vec{p}_a + \vec{p}_b = 0$ and $E_a + E_b = \sqrt{s}$, so the above expression can be simplified to
\begin{equation}
  F = s - 2\sqrt{s}(E_1+E_3) + m_1^2 + m_3^2 + 2p_1 p_3 - m_2^2.
\end{equation}

Now we are going to express it in terms of new variables:
\begin{align}
  F &= s - 2\sqrt{s}(m_{1,t}\cosh{y_1} + m_{3,t}\cosh{y_3}) + 2m_{1,t}m_{3,t}(\cosh{y_1} \cosh{y_3} - \sinh{y_1} \sinh{y_3}) \nonumber \\
  & + 2p_{1,t}p_{3,t}\cos\varphi + m_1^2 + m_3^2 - m_2^2
  = s - 2\sqrt{s}(\sqrt{p_{1,t}^2+m_1^2+p_{1,z}^2} + E_3) \nonumber \\
  & + 2E_3\sqrt{p_{1,t}^2+m_1^2+p_{1,z}^2} -2p_{1,z}p_{3,z} 
  + 2p_{1,t}p_{3,t}\cos\varphi + m_1^2 + m_3^2 - m_2^2.
\end{align}

The derivative of $F$ reads
\begin{align}
  \frac{\partial F}{\partial y_1} &= -2\sqrt{s}(m_{1,t}\sinh{y_1}) + 2m_{1,t}m_{3,t}(\sinh{y_1}\cosh{y_3} - \cosh{y_1}\sinh{y_3}) \nonumber \\
  & = - 2\left[ p_{1,z}(\sqrt{s}-m_{3,t}\cosh{y_3}) + \sqrt{p_{1,t}^2+m_1^2+p_{1,z}^2} m_{3,t} \sinh{y_3} \right].
\end{align}

The solutions for $F=0$ are
\begin{equation}
  p_{1,z} = \frac{-b \pm \sqrt{b^2 - 4ac}}{2a},
\end{equation}
with
\begin{align}
  a &= (\sqrt{s} - E_3)^2 - p_{3,z}^2, \nonumber \\
  b &= p_{3,z} \, \eta, \nonumber \\
  c &= (\sqrt{s}-E_3)^2 (p_{1,t}^2+m_1^2) - \frac{1}{4} \eta^2, \nonumber \\
  \eta &= s - 2\sqrt{s}E_3 + 2p_{1,t}p_{3,t}\cos\varphi + m_1^2 + m_3^2 - m_2^2.
\end{align}
The solutions have to satisfy the additional constraint
\begin{equation}
  s - \sqrt{s}E_3 + 2p_{1,t}p_{3,t}\cos{\varphi} - 2p_{1,z}p_{3,z} + m_1^2 + m_3^2 - m_2^2 \geq 0.
\end{equation}
By performing the analytical integration over the $\delta$ function, we obtain the final expression for the phase space:
\begin{equation}
  d\Phi_3 = \frac{1}{128 \pi^4} dp_{1,t}^2 dp_{3,t}^2 dy d\varphi \sum\left( \left.\frac{\partial F}{\partial y_1}\right|_{F=0} \right)^{-1},
\end{equation}
and the total cross section reads
\begin{equation}
  \sigma_{2\to 3} = \frac{1}{128 \pi^4 F} \int_{0}^{p_C} dp_{1,t}^2 \int_0^{p_C} dp_{3,t}^2 \int_{y_{\min}}^{y_{\max}} dy \int_0^\pi d\varphi \Msq{2\to 3} \sum\left( \left.\frac{\partial F}{\partial y_1}\right|_{F=0} \right)^{-1},
\end{equation}
where $p_C = \lambda(\sqrt{s},m_1,m_2)/(2\sqrt{s})$ and $y_{\max/\min}=\pm \text{arccosh}(\frac{\sqrt{p_C^2+m_3^2}}{m_{3,t}})$.

\end{widetext}


\bibliography{refs}
\end{document}